\documentclass[aps,pra,twocolumn,superscriptaddress,showpacs]{revtex4-1}
\usepackage{amsmath,amssymb,amsfonts,amsthm}
\usepackage{graphicx}
\usepackage{bbm}
\usepackage{dcolumn}   
\usepackage{epstopdf}
\usepackage[caption=false]{subfig}
\usepackage{color}
\usepackage[colorlinks=true,linkcolor=blue,urlcolor=blue,citecolor=blue]{hyperref}

\begin{document}

 \newcommand{\breite}{1.0} 

\newtheorem{prop}{Proposition}
\newtheorem{cor}{Corollary}

\newcommand{\be}{\begin{equation}}
\newcommand{\ee}{\end{equation}}

\newcommand{\bea}{\begin{eqnarray}}
\newcommand{\eea}{\end{eqnarray}}
\newcommand{\lt}{<}
\newcommand{\gt}{>}

\newcommand{\Reals}{\mathbb{R}}     
\newcommand{\Com}{\mathbb{C}}       
\newcommand{\Nat}{\mathbb{N}}       

\newcommand{\ic}[1]{{\color{red} #1}}

\newcommand{\id}{\mathbbm{1}}    

\newcommand{\Real}{\mathop{\mathrm{Re}}}
\newcommand{\Imag}{\mathop{\mathrm{Im}}}

\def\O{\mbox{$\mathcal{O}$}}   
\def\F{\mathcal{F}}			
\def\sgn{\text{sgn}}

\newcommand{\deo}{\ensuremath{\Delta_0}}
\newcommand{\dea}{\ensuremath{\Delta}}
\newcommand{\ak}{\ensuremath{a_k}}
\newcommand{\ad}{\ensuremath{a^{\dagger}_{-k}}}
\newcommand{\sx}{\ensuremath{\sigma_x}}
\newcommand{\sz}{\ensuremath{\sigma_z}}
\newcommand{\spl}{\ensuremath{\sigma_{+}}}
\newcommand{\smi}{\ensuremath{\sigma_{-}}}
\newcommand{\alk}{\ensuremath{\alpha_{k}}}
\newcommand{\bk}{\ensuremath{\beta_{k}}}
\newcommand{\ok}{\ensuremath{\omega_{k}}}
\newcommand{\vd}{\ensuremath{V^{\dagger}_1}}
\newcommand{\vi}{\ensuremath{V_1}}
\newcommand{\vo}{\ensuremath{V_o}}
\newcommand{\zc}{\ensuremath{\frac{E_z}{E}}}
\newcommand{\xc}{\ensuremath{\frac{\Delta}{E}}}
\newcommand{\xd}{\ensuremath{X^{\dagger}}}
\newcommand{\aok}{\ensuremath{\frac{\alk}{\ok}}}
\newcommand{\tpw}{\ensuremath{e^{i \ok s }}}
\newcommand{\tpe}{\ensuremath{e^{2iE s }}}
\newcommand{\tmw}{\ensuremath{e^{-i \ok s }}}
\newcommand{\tme}{\ensuremath{e^{-2iE s }}}
\newcommand{\epls}{\ensuremath{e^{F(s)}}}
\newcommand{\emis}{\ensuremath{e^{-F(s)}}}
\newcommand{\epl}{\ensuremath{e^{F(0)}}}
\newcommand{\emi}{\ensuremath{e^{F(0)}}}

\newcommand{\lr}[1]{\left( #1 \right)}
\newcommand{\lrs}[1]{\left( #1 \right)^2}
\newcommand{\lrb}[1]{\left< #1\right>}
\newcommand{\nbt}{\ensuremath{\lr{ \lr{n_k + 1} \tmw + n_k \tpw  }}}

\newcommand{\om}{\ensuremath{\omega}}
\newcommand{\dw}{\ensuremath{\Delta_0}}
\newcommand{\wbp}{\ensuremath{\omega_0}}
\newcommand{\dv}{\ensuremath{\Delta_0}}
\newcommand{\vbp}{\ensuremath{\nu_0}}
\newcommand{\vplus}{\ensuremath{\nu_{+}}}
\newcommand{\vminus}{\ensuremath{\nu_{-}}}
\newcommand{\wplus}{\ensuremath{\omega_{+}}}
\newcommand{\wminus}{\ensuremath{\omega_{-}}}
\newcommand{\uv}[1]{\ensuremath{\mathbf{\hat{#1}}}} 
\newcommand{\abs}[1]{\left| #1 \right|} 
\newcommand{\avg}[1]{\left< #1 \right>} 
\let\underdot=\d 
\renewcommand{\d}[2]{\frac{d #1}{d #2}} 
\newcommand{\dd}[2]{\frac{d^2 #1}{d #2^2}} 
\newcommand{\pd}[2]{\frac{\partial #1}{\partial #2}} 
\newcommand{\pdd}[2]{\frac{\partial^2 #1}{\partial #2^2}} 
\newcommand{\pdc}[3]{\left( \frac{\partial #1}{\partial #2}
 \right)_{#3}} 
\newcommand{\ket}[1]{\left| #1 \right>} 
\newcommand{\bra}[1]{\left< #1 \right|} 
\newcommand{\braket}[2]{\left< #1 \vphantom{#2} \right|
 \left. #2 \vphantom{#1} \right>} 
\newcommand{\matrixel}[3]{\left< #1 \vphantom{#2#3} \right|
 #2 \left| #3 \vphantom{#1#2} \right>} 
\newcommand{\grad}[1]{\gv{\nabla} #1} 
\let\divsymb=\div 
\renewcommand{\div}[1]{\gv{\nabla} \cdot #1} 
\newcommand{\curl}[1]{\gv{\nabla} \times #1} 
\let\baraccent=\= 

\title{$1/f^\alpha$ noise and generalized diffusion in random Heisenberg spin systems}

\author{Kartiek Agarwal}
\affiliation{Physics Department, Harvard University, Cambridge, Massachusetts 02138, USA}
\email[]{agarwal@physics.harvard.edu}
\author{Eugene Demler}
\affiliation{Physics Department, Harvard University, Cambridge, Massachusetts 02138, USA}
\author{Ivar Martin}
\affiliation{Material Science Division, Argonne National Laboratory, Argonne, Illinois 60439USA}

\date{\today}
\begin{abstract}

We study the `flux noise' spectrum of random-bond quantum Heisenberg spin systems using a real-space renormalization group (RSRG) procedure that accounts for both the renormalization of the system Hamiltonian and of a generic probe that measures the noise. For spin chains, we find that the dynamical structure factor $S_q(f)$, at finite wave-vector $q$, exhibits a power-law behavior both at high and low frequencies $f$, with exponents that are connected to one another and to an anomalous dynamical exponent through relations that differ at $T = 0$ and $T = \infty$. The low-frequency power-law behavior of the structure factor is inherited by any generic probe with a finite band-width and is of the form $1/f^\alpha$ with $0.5 < \alpha < 1$. An analytical calculation of the structure factor, assuming a limiting distribution of the RG flow parameters (spin size, length, bond strength) confirms numerical findings. More generally, we demonstrate that this form of the structure factor, at high temperatures, is a manifestation of anomalous diffusion which directly follows from a generalized spin-diffusion propagator. We also argue that $1/f$-noise is intimately connected to many-body-localization at finite temperatures. In two dimensions, the RG procedure is less reliable; however, it becomes convergent for quasi-one-dimensional geometries where we find that one-dimensional $1/f^\alpha$ behavior is recovered at low frequencies; the latter configurations are likely representative of paramagnetic spin networks that produce $1/f^\alpha$ noise in SQUIDs.

\end{abstract}
\maketitle

\section{Introduction} 
Disordered interacting spin systems display a variety of physical phenomena: the spin glass transition~\cite{edwards1975theory,fisherdroplet,kirkpatrickinfrange,macmillanscaling,mezard1987spinglassbook} and slow relaxation~\cite{randeriaspinglasslowf,fisherdynamics,Yu,Ghoshspinosccilations}, many-body localization transition~\cite{palmbl2010,sarangmblspectral,mooremblentangle,sebrynlocal,oganesyan2007} and a breakdown of ergodicity, Griffiths (rare-region) effects~\cite{agarwal2014anomalous,qgriffithshuse,griffithsheisenberg,qgriffithsvojta}, strong-randomness fixed points~\cite{motrunichdamlehuse,mortrunichising,thefisher}, spin-liquid states~\cite{sachdevgapless} and even excited topological states~\cite{bahri2013localization}. A great deal of our understanding of these phenomena has come from the development of the real-space renormalization group (RSRG) method~\cite{dasguptama,thefisher92}. Previous applications of the method focussed on studying the low-temperature thermodynamic properties of disordered spin chains. More recently, the approach has been applied to study both high temperature and dynamical properties---this is based on the identification that the protocol, more generally, involves the creation of local integrals of motion, i.e., the high-energy modes eliminated during the RG process are pieces of approximate many-body eigenstates of the system~\cite{pekker,motrunichdamlehuse,vasseur2014dynamics}. Using such ideas, theoretical analysis of the low-temperature optical conductivity of Anti-Ferromagnetic spin chains of various kinds was carried out in Ref.~\cite{motrunichdamlehuse} while that for Ising spin chains was performed computationally to identify a transition between various infinite-temperature many-body localized phases in Ref.~\cite{pekker}. 

\begin{center}
\begin{figure}[t]
\includegraphics[width = 3.5in]{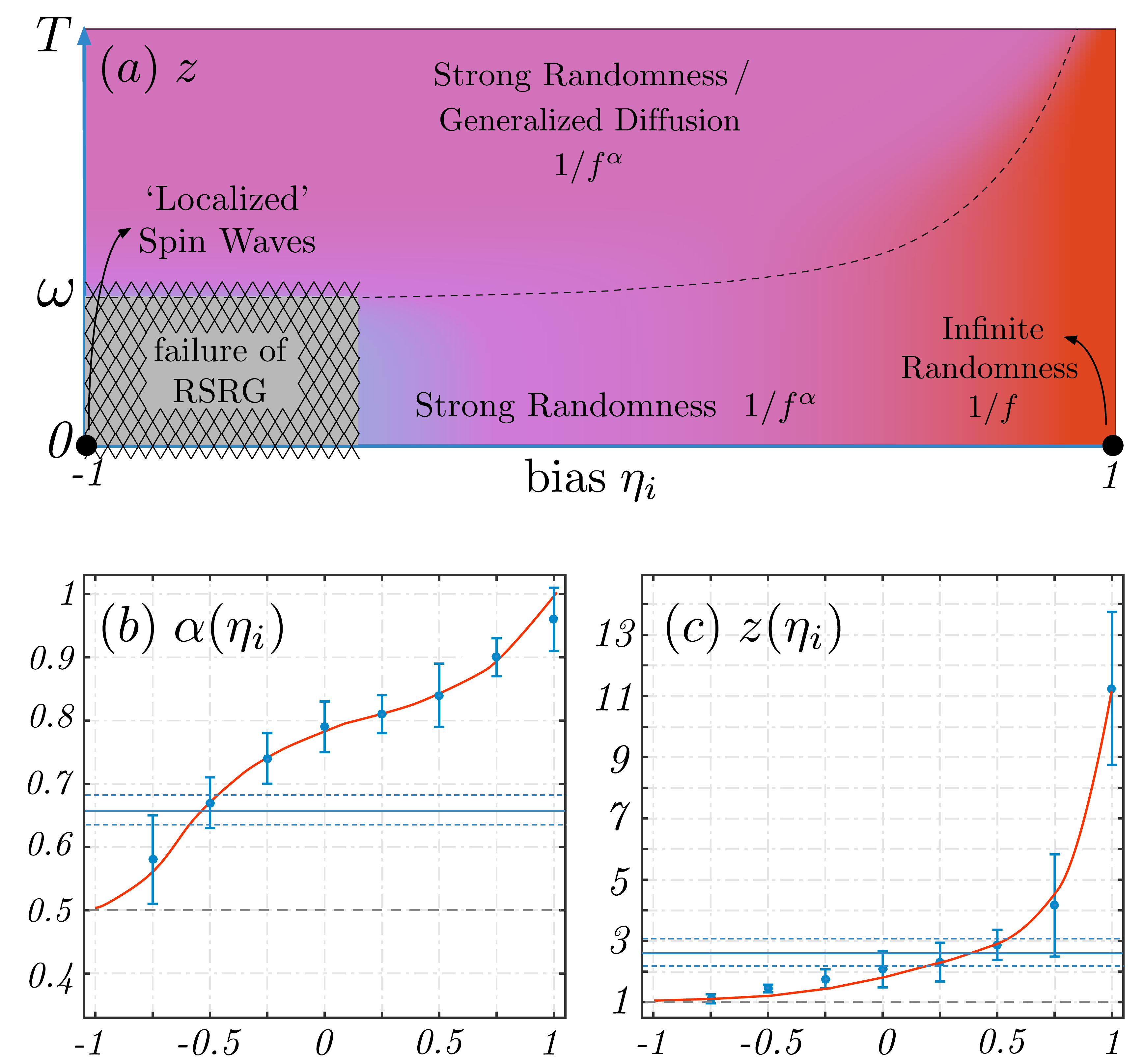}
\caption{(a) Phase diagram of the \emph{strongly} disordered Heisenberg spin chain (with $1/|J|$ distribution of initial couplings in a range $|J| \in [1,e^D]$ ) extrapolated to finite temperatures using RSRG results at zero and infinite temperatures;  the color scheme encodes the variation of the dynamical exponent $z$ as a function of the bias $\eta_i$ [initial proportion of F/AF (1/-1) bonds in chain]. (b) The low-frequency noise power law $\alpha$ and (d) the dynamical critical exponent $z$ are plotted against $\eta_i$ at $T = 0$ (data-points and error bars) and $T = \infty$ (flat line with dashed lines indicating error), for disorder strength $D = 3$ (see main text). At $T = 0$, the purely F system harbors heavily damped spin wave excitations whose localization length diverges as $1/\omega$ in the zero-frequency limit. The purely AF system, which is known to flow to a infinite-randomness fixed point, exhibits $1/\omega$ noise. Intermediate biases (non-hatched region) flow to strong-randomness fixed points accompanied by $1/\omega^\alpha$ noise spectra. Hatched region is inaccessible to RSRG. At high temperatures ($T \gg \omega$), the $1/\omega^\alpha$ noise spectra is generic to \emph{strongly} disordered Heisenberg chains and can be interpreted in the context of generalized diffusion. The crossover from zero-temperature to infinite-temperature behavior occurs at $T / \omega \sim 2^{z_0}$, where $z_0$ is the dynamical exponent found for the zero temperature system.}
\label{fig:figalphaz}
\end{figure}
\end{center}

 The work presented here builds on and extends the RSRG program to address a crucial question in the dynamics of disordered systems---how do they generate scale-invariant $1/\omega^\alpha$ (or $1/f^\alpha$) noise? Numerous experiments~\cite{Wellstood,Clarke2,Wellstoodtemp,Martinis,Siddiqi,Sendelbach,Oliver} on Superconducting Quantum Interference devices (SQUIDs) observe a flux noise with a spectrum $N(\omega) \sim 1/\omega^\alpha$ whose magnitude is nearly temperature independent, but exponent $\alpha$ changes smoothly with temperature~\cite{Clarke2,Wellstoodtemp}. The noise likely originates from fluctuating electronic spins localized on the metal-insulator (the conducting strip and the substrate) interface of the SQUIDs~\cite{Wellstood,Koch,Ioffe}. Such spins can interact via oscillatory RKKY (Heisenberg) interactions~\cite{Ioffe} whose sign flips on the order of the Fermi-wavelength and is effectively random at the scale of separation of the spins, i.e., interactions can be both Ferromagnetic (F) or Anti-Ferromagntic (AF). The dynamics of such spins, and how they generate the observed noise spectrum is less understood. In particular, it was posited that the two-dimensional surface spins exhibit regular diffusion~\cite{Ioffe,rogeriodwave} and that the $1/\omega$ noise spectrum arises in a limited frequency range owing to \emph{specifics} of the geometry of the probe coupling. Such explanations predict a large frequency lower bound to the $1/\omega$ form of the noise spectrum which is not observed in experiments. More importantly, it was assumed that the disorder is self-averaging (which leads to diffusion). A central finding of our work is that, for sufficiently strong initial disorder, this is not the correct conclusion for one-dimensional and two-dimensional `strips' of Heisenberg spins---we find that such interacting spin networks generically flow towards strong-randomness fixed points where they exhibit anomalous diffusion and, moreover, this entails a flux noise that \emph{intrinsically} exhibits a $1/\omega^\alpha$ ($\alpha < 1$) noise spectrum. 
  
The RSRG protocol we develop allows us to directly compute the noise spectrum $N(\omega)$ measured by a probe that couples (through arbitrary geometrical factors) to the flux generated by spins interacting via disordered exchange couplings, at both high and low temperatures. Our approach is elaborated upon in Sec.~\ref{sec:RSRGrules}. A numerical application of the protocol suggests that a (strongly) disordered Heisenberg spin chain flows towards a strong-randomness fixed point characterized by an anomalous dynamical exponent $z = 1/\beta \neq 1$,$2$. Moreover, this dynamical exponent determines the form of the noise; a harmonic probe with wave-vector $q$ measures the noise $S_q(\omega)$ (the dynamical structure factor) which shows a piece-wise power-law frequency dependence: $S_q (\omega) \sim 1/\omega^{\alpha}$ for $\omega \ll q^{1/\beta}$ and $S_q(\omega) \sim 1/\omega^{\alpha'}$ for $\omega \gg q^{1/\beta}$. The precise values of these power laws are independent of the initial composition (proportion of F/AF bonds) of the spin chain \emph{if} the distribution of couplings is sufficiently non-singular, but varies for more singular initial distributions such as those relevant for experiments on SQUIDs (and that we primarily consider)---for such distributions, the low-frequency noise exponent varies in the range $0.5 < \alpha < 1$ (Fig.~\ref{fig:figalphaz}). [Note that any probe of finite spatial bandwidth will inherit the low frequency exponent $\alpha$ in the structure factor $S_q (\omega)$.] Our numerical results for spin chains are discussed in Sec.~\ref{sec:numsimulations}. 

While exponents $z$, $\alpha$, and $\alpha'$ can take a range of values, they are tied by the non-trivial relations $\alpha' = 1 + 2(1 - \alpha)$, $\alpha = 1-1/z$ at high temperatures, and $\alpha' = 1 + 3(1-\alpha)$, $\alpha = 1-1/2z$ at low temperatures. In Sec.~\ref{sec:scalingapproach}, we show how these relations can be obtained from a scaling form of the probability distribution function $P_{F (AF)}$ that governs the distribution of F(AF) bonds with a given coupling at energy scales where the RG procedure has converged. These relations are then verified numerically by performing a scaling collapse of $S_q (\omega)$ for a wide range of wave-vectors and frequencies (Fig. \ref{fig:figmain}) at both $T = 0$, and $T = \infty$; the success of the scaling collapse gives further credence to the validity of the RG procedure. 

While the above discussion pertains specifically to the Heisenberg spin chain, we show, in Sec.~\ref{sec:generalizedapproachandmbl}, that the form of the dynamical structure factor carries over generally, at high temperatures, independently of dimensionality, to systems exhibiting anomalous diffusion. We propose a generalized-diffusion ansatz for the spin propagator, $G_q (\omega) = 1/[-i \omega + q^{1/\beta} f(\omega/q^{1/\beta})]$, that can account for anomalous diffusion and show that it directly reproduces the limiting $q-$ and $\omega-$dependencies of the structure factor. Thus, we conclude that at high temperatures ($T \gg \omega$), $1/\omega^\alpha$ noise is generic to a system exhibiting anomalous diffusion (accompanied by an anomalous dynamical exponent). The generalized-diffusion approach, however, fails to explain the structure factor at $T = 0$. We discuss how this is a consequence of the failure of linear response at $T=0$---we show that the system exhibits a divergent static susceptibility for even finite-$q$ perturbations. 

The $1/\omega^\alpha$ low-frequency behavior of the structure factor also has the consequence that all (harmonic) spin correlations decay to zero in the long time limit as $1/t^{1-\alpha}$. Conversely, when $\alpha = 1$ precisely, even finite-$q$ modes, which are not conserved are unable to equilibrate, signaling the onset of the many-body localization (MBL) transition. Thus, $1/\omega$ noise is connected to the MBL transition. A hint of this is observed in our simulations at $T = \infty$, where, upon increasing the disorder strength, the low-frequency noise exponent approaches $1$, but does not reach it. At $T = 0$, the noise exponent $1$ is obtained in the case of a purely AF chain, where the RG is expected to flow towards \emph{infinite}-randomness fixed point~\cite{dasguptama,fisheraf}. The status of ergodicity in our model is less clear. The RG constructs a macroscopic number of integrals of motion which suggests a lack of ergodicity. However, these integrals of motion are not exact since the RG flows towards a strong-randomness fixed point, as opposed to infinite-randomness where it does become probabilistically exact. We anticipate future studies should elucidate this question. (While MBL implies a lack of ergodicity, the converse is not true; see for instance Ref.~\cite{ioffe_nonergodic}.)

In Sec.~\ref{sec:2dsimulsmain} we discuss the application of the RSRG procedure to two-dimensional spin networks. For square geometries, at high temperatures, the structure factor shows scaling behavior (although in a limited range of frequencies) according to the generalized-diffusion ansatz for a dynamical exponent $z \approx 2$. Note that regular diffusion is also associated with $z = 2$; however, in the generalized-diffusion framework, this implies that the finite-$q$ structure factor has a $1/\sqrt{\omega}$ low-frequency tail instead of the flat spectrum expected in the case of regular diffusion. We note that for square geometries, the accuracy of the decimation procedure does not improve significantly over the course of the RG. Therefore, we cannot rule out (or confirm) regular diffusion. In contrast, for an elongated two-dimensional `strip', we find that at the lowest frequencies, this $1/\sqrt{\omega}$ tail transitions into an anomalous $1/\omega^\alpha$ tail ($ 0.5 < \alpha < 1$) as in the one-dimensional case; thus, as the clusters become wider than the width of the two-dimensional strip, their collective dynamics mimics that of a one-dimensional array of disordered spins. We end by summarizing our results and their connection to experiments in Sec.~\ref{sec:summary} and proposing further possible extensions of our RSRG protocol to systems with anisotropic couplings. 

\begin{center}
\begin{figure*}[t]
\includegraphics[width = 7 in]{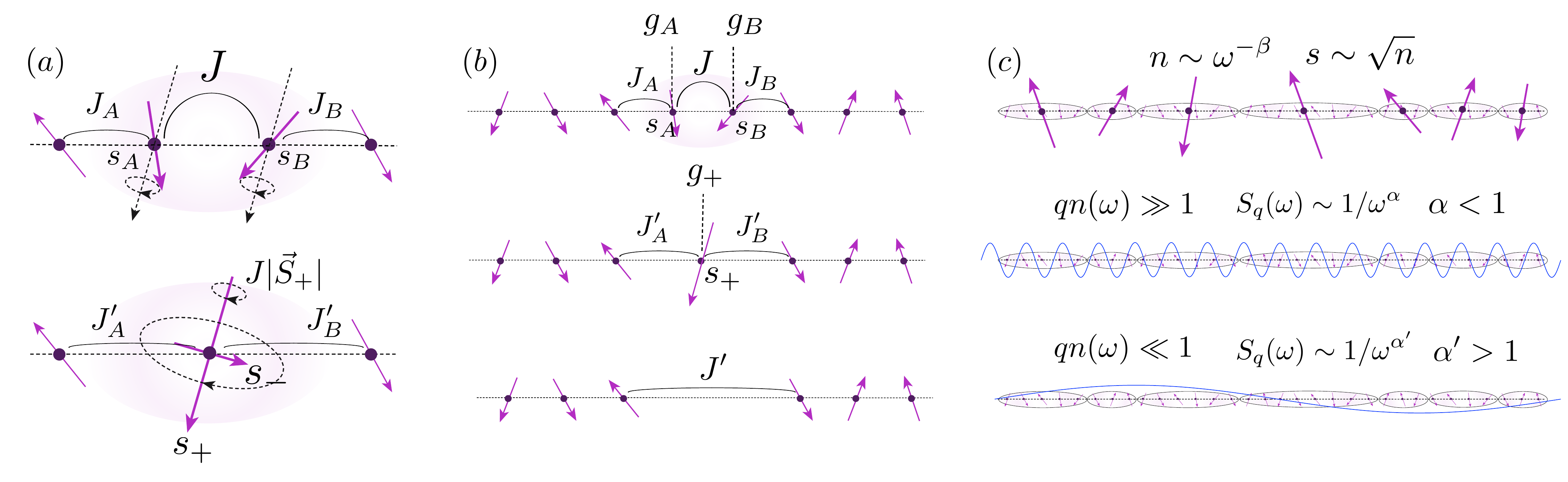}
\caption{(a) Two-spin dynamics. Two spins $A$ and $B$ strongly coupled to one another revolve around their net moment $\vec{S}_+ = \vec{S}_A + \vec{S}_B$ with a large frequency $J |\vec{S}_+|$, while $\vec{S}_+$ moves slowly owing to weak interactions $J'_A$ and $J'_B$. $\vec{S}_- = \vec{S}_A - \vec{S}_B$ is effectively decoupled from the slow dynamics of $\vec{S}_+$ and results in the `noise' evaluated at this RG step. (b) Demonstration of RG rules. The strongly coupled spin-pair are decimated to form an effective spin (with quantized net moment $s_+$), and Heisenberg couplings are renormalized from $J_{A (B)}$ to $J_{A'(B')}$. The effective spin couples to the probe with amplitude $g_+$. If a singlet is formed, both spins are integrated out, providing an effective interaction $J'$ between the neighboring spins. (c) Scaling behavior at $T = \infty$ (at $T = 0$, the bonds, composed of long-range singlets scale as the cluster size). Size of clusters $n$, their spin size $s$ and the time-scale of their dynamics $1/\omega$ are connected through the anomalous dynamical exponent $1/\beta$. Harmonic probes (illustrated in blue) measure a low-(high-)frequency noise-exponent $\alpha$ or $\alpha'$ in the regimes where the probe length $1/q$ is smaller (greater) than the clusters whose dynamics they probe.}
\label{fig:figrules}
\end{figure*}
\end{center}
\section{RSRG approach for computing noise} 
\label{sec:RSRGrules}
The model we study is the disordered Heisenberg spin-1/2 system with Hamiltonian $H = \sum_{ij} J_{ij} \vec{S}_i .\vec{S}_j$. $J_{ij}$'s are picked independently from a distribution $P_0(J)$ with support over both negative and positive values of $J$. The probe measures a magnetic flux $\vec{M} = \sum_{i} g_i \vec{S}_i$, and the noise measured by such a probe has a spectral form $N(\omega) = \mathcal{F} \{ \avg{[\vec{M}(t).,\vec{M}(0)]_+} \}$, that is, the Fourier transform of the autocorrelation of $\vec{M}(t)$. (Such an autocorrelation directly yields the $x-,y-,z-$component-wise noise auto-correlations due to spin-rotation symmetry of the model. Also note that the methodology presented here can be straightforwardly extended to the case where $g_i$s are vectors measuring a certain the projection of the local spin, which can be non-uniform. We do not treat this general case here for the sake of simplicity.) The RSRG approach and the evaluation of the noise spectrum can be summarized succinctly as follows. We first find the 2 spins $\vec{S}_A$ and $\vec{S}_B$ that are most strongly coupled to one another; these spins tend to precess rapidly around their total moment $\vec{S}_+ = \vec{S}_A + \vec{S}_B$, which on the other hand, moves slowly if the constituent spins are relatively weakly coupled to the external neighbors. The total spin $\vec{S}_+$ is then kept as an effective spin with an effective interaction with the probe and its neighbors, while the rapid precession of the constituent spins is counted towards the `noise' at the frequency at which they precess about $\vec{S}_+$. This procedure is repeated ad-infinitum. The approach is illustrated in Figs.~\ref{fig:figrules} (a) and (b).

It must be noted that the quantum-mechnical derivation of the ground state RSRG rules for bond decimation in mixed AF/FM case for was first carried out in Ref.~\cite{westerberg}. However, as mentioned above, to calculate the noise spectrum at arbitrary temperature, we need to extend these rules to arbitrary composite states and provide renomalization rules for the probe function  $g_i$. In order to keep the discussion self-contained and shed light on the physics of the problem, we discuss the derivation of these rules using a straightforward semi-classical approach, that yields the quantum-mechnical results upon spin re-quantization.

\subsection{Derivation of RG rules}
To obtain the general bond decimation rules, let us first consider a set of two strongly interacting spins $\vec{S}_A$ and $\vec{S}_B$ with mutual coupling $J$ and couplings $J_A$ and $J_B$ to external spins, such that $J \gg J_A, J_B$. Due to their large mutual coupling $J$, spins A and B precess about their combined moment $\vec{S}_+ = \vec{S}_A + \vec{S}_B$ with a large frequency $\Delta_0= J \abs{\vec{S}_+}$. The neighboring spins couple only to the slow motion of spins $\vec{S}_A$ and $\vec{S}_B$ which is the projection of these spins on to the slow combined moment $\vec{S}_+$. Thus, the effective couplings $J_{A(B)}$ are modified to $J'_{A (B)} = J_{A (B)} ( \vec{S}_{A (B)} \cdot\vec{S}_+  )/ \abs{\vec{S}_+}^2$. The quantum-mechanical result is obtained by re-quantizing the spins, i.e., $|\vec{S}_A|^2 = s_A (s_A + 1)$,  $|\vec{S}_B|^2 = s_B (s_B + 1)$, and  $|\vec{S}_+|^2 = s (s + 1)$. 

The quantum-mechanical interpretation of the above procedure is that, at every step of the RG, we solve the approximate Hamiltonian $H_{AB} = J \vec{S}_A. \vec{S}_B + ...$ to first order in perturbation theory by projecting the system to a particular eigenstate of $\vec{S}_+$ with quantum number $s$. This effective spin has couplings $J'_A$ and $J'_B$ determined above with its neighbors. The frequency $\Delta_0$ with which the spins precess about the total moment $\vec{S}_+$ is split into 2 in the quantum-mechanical situation, and is given by the energy difference between total angular momentum states $s$ and $s \pm 1$ of the coupled spins. The perturbation theory is controlled by the parameter $\Delta_{A(B)}/\Delta_0$ where $\Delta_{A}$,$\Delta_B$ correspond to a similar gaps generated by the combination of spins $A$ and $B$ with their neighbors. Note that in the case of singlet-formation ($s = 0$), second-order perturbation theory must be performed, eliminating spins $A$ and $B$ altogether and yielding an effective coupling between their neighboring spins, $J' = 2 s_A (s_A + 1) J_A J_B / 3 J$ ($s_A = s_B$)~\cite{westerberg}. Note that in higher dimensions, when external spin $A$ (and/or $B$) may couple to both spins $1$ and $2$ undergoing a singlet formation with couplings $J_{A,1}$ and $J_{A,2}$ respectively, the coupling $J_A$ in this result should be replaced by $J_{A,1} - J_{A,2}$.

The renormalization of the probe coupling is also immediate: if the probe couples with strength $g_A$ and $g_B$ to spins $A$ and $B$, it couples to the effective spin $\vec{S}_+$ with a strength $ g_+ = g_A  ( \vec{S}_{A} \cdot \vec{S}_+  )/ \abs{\vec{S}_+}^2 + g_B  ( \vec{S}_{B} \cdot \vec{S}_+  )/ \abs{\vec{S}_+}^2 = (g_A + g_B)/2 +  (g_A - g_B)/2 \left[\abs{\vec{S}_A}^2 -\abs{\vec{S}_B}^2 \right] / \abs{\vec{S}_+}^2$. Note that in the case of singlet formation, the spins are eliminated altogether and the probe does not couple to the singlet. This will result in important differences in the behavior of the system at $T = 0$ (where singlets are preferred) and $T = \infty$ (where singlets are unfavorable for entropic reasons).

\subsection{Evaluation of Noise at every RG step}
We want to evaluate the dynamics of the object $M(t) = \sum_i g_i \vec{S}_i $. As mentioned above, we do this in a step-by-step RG fashion. In particular, the flux can be partitioned into a slow part $M_S(t) = \sum_{i \neq A,B} g_i \vec{S}_i + g_+ \vec{S}_+$ whose dynamics is determined in subsequent RG steps, and the remaining fast part $M_F(t) = (g_A - g_B) / 2 \left[ \vec{S}_- - \vec{S}_+(\vec{S}_- \cdot \vec{S}_+) / \abs{\vec{S}_+}^2 \right]$, where $\vec{S}_- = \vec{S}_A - \vec{S}_B$ and the combination in square brackets is the component  of $\vec{S}_-$ orthogonal to $\vec{S}_+$. To zeroth order in perturbation theory (in $\Delta_{A(B)}/\Delta$), the noise spectrum $N(\omega)$ receives a contribution $\mathcal{F} \{ \avg{[M_F(t),M_F(0)]_+} \}$ from this step of the RG, where the brackets $\avg{}$ correspond to a quantum mechanical expectation in an eigenstate $\ket{s,m}$ with $|\vec{S}_+|^2 = s ( s + 1) $ and an arbitrary projection quantum number  $m$ (there is no preferred axis for the total moment $\vec{S}_+$). 

Note that, $\vec{S}_-$ produces transitions from states $\ket{s,m}$ to states $\ket{s\pm 1, m \pm 1}$. Due to the full rotational symmetry of the problem, the transition rates are independent of $m$, and the transition frequencies only depend on $s$. Hence, the noise associated with these contributions is $N(\omega) = (g_A - g_B)^2/4 \left[ M(s,\uparrow) \delta ( \omega - \omega_\uparrow) + M(s,\downarrow) \delta ( \omega + \omega_\downarrow) \right]$, where $\omega_{\uparrow (\downarrow)}$ is the frequency of the transition from the angular momentum state $s$ to $s+1$ ($s$ to $s-1$) and $M(s, \uparrow (\downarrow))$ is the accompanying matrix element of this transition. In what follows, we will refer to the factor $(g_A - g_B)^2/4$ as the `probe form factor' since this factor depends on the precise details of the probe, and the matrix elements $M(s, \uparrow (\downarrow))$ corresponding to the spin operators will be referred to as the `noise amplitude'. The precise evaluation of the noise amplitude is discussed in Appendix~\ref{sec:noisecalc}. 

\section{Numerical Simulations in One Dimension} 
\label{sec:numsimulations}
\subsection{Implementation}

\begin{center}
\begin{figure}[t]
\includegraphics[width = 3in]{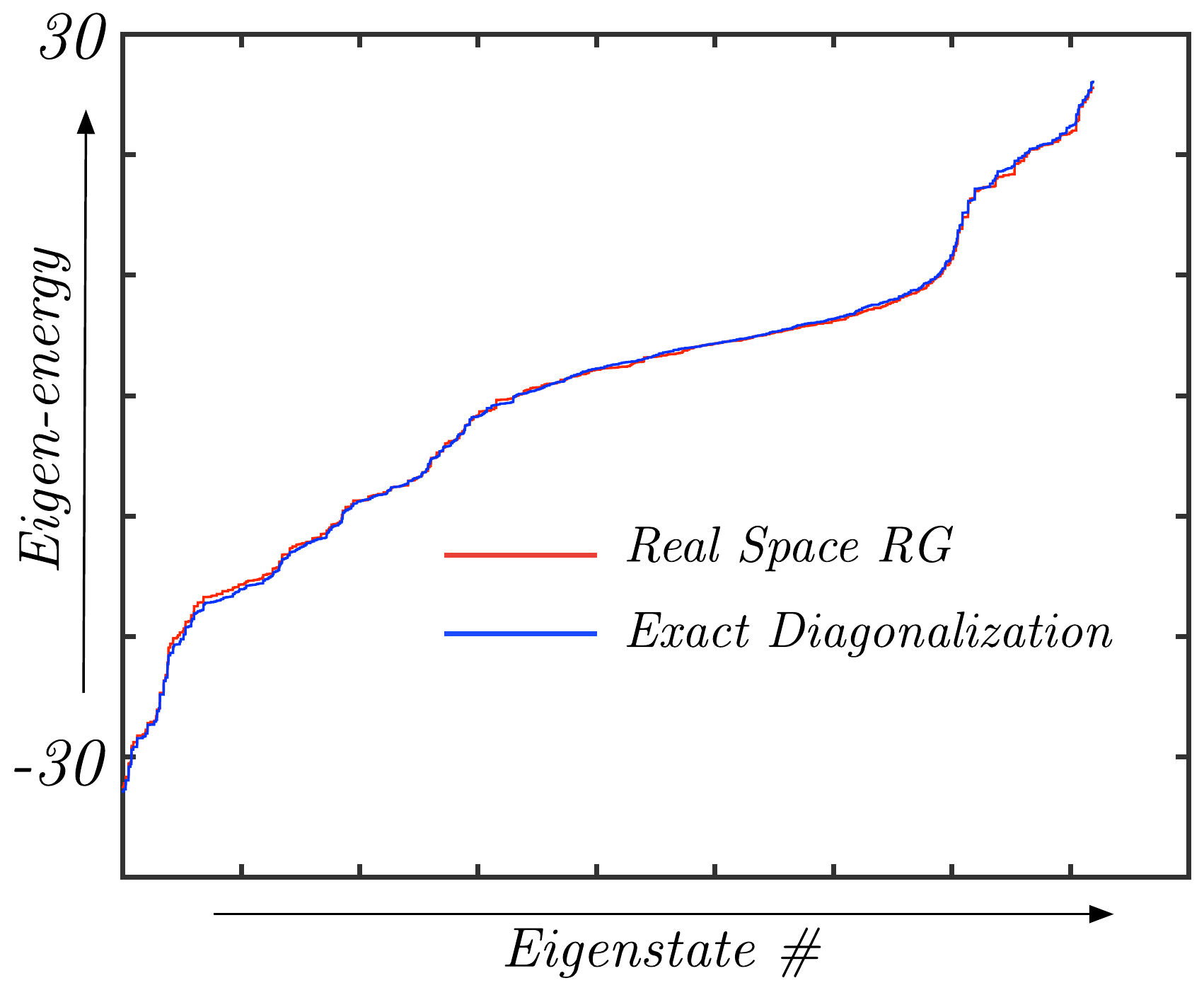}
\caption{RG-determined spectrum (red) compared with exact digaonlization determined spectrum (blue) for a particular $12$-site random spin chain with initial $1/|J|$ distribution, disorder strength $D = 3$, and equal mix of F/AF bonds.}
\label{fig:rged}
\end{figure}
\end{center}

As a first check of the efficacy of the RG rules, we perform a direct comparison of the complete energy spectrum determined by exact-diagonalization and the RG procedure for small random spin chains. A typical run is shown is Fig.~\ref{fig:rged}; the RG result appears to be in good agreement with the exact-diagonaliation result. (More quantitative checks of the RG procedure are discussed in Appendix~\ref{sec:numchecks}.) For the determination of the noise and the structure factor, we perform numerical simulations for spin chains of size $L = 15000$ at two fixed temperatures $T = 0$ and $T = \infty$. The majority of our simulations assume distribution of magnitude of the initial couplings $P_0(|J|)$ is of the form $P_0(|J|) \sim 1/|J|$, with values in the range $|J| \in [ 1, e^D]$. This choice is motivated by the observation that a system of spins scattered randomly in $d$ dimensions interacting with a $1/r^3$ RKKY interaction, has a distribution $P_0(|J|) \sim 1/|J|^{1 + d/3}$; the additional factor of $d/3$ is not systematically considered since it is found to not affect the results qualitatively. (In Ref.~\cite{westerberg}, Westerberg et al. work with an array of initial distributions and conclude that distributions more singular than $P_0(|J|) \sim 1/|J|^{0.7}$ flow to non-universal fixed points at $T = 0$.) Another simplification we make is to consider interactions only between nearest-neighbor spins even though spins interact with all other spins via the RKKY mechanism; this is well justified because $1/r^3$ interactions are sufficiently short-ranged in one and quasi-one dimensional cases we primarily consider. We also perform simulations with a uniform distribution (range $|J| \in [ 0 , 1]$). We choose a finite proportion of these couplings to be AF(F) and characterize this initial `bias' by a variable $\eta_i \in (-1, 1)$: $\eta_i = +1 (-1) $ corresponds to a system with purely AF(F) couplings. To perform the zero temperature calculations, at every RG step, we choose the effective spin to reside in a state with $s = | s_A - s_B | $ ( $ s = s_A + s_B$ ) for spins $A$ and $B$ that are coupled by a AF(F) bond. At infinite temperature, the quantum number $s$ is chosen probabilistically according to the degeneracy $2s + 1$ associated with the state.  

The choice of the probe is another free parameter to be considered in our problem. The most natural choice of probe is $g_i = \cos( q i )$, that is, a harmonic probe that measures the dynamical structure factor $S_q (\omega)$ at a given wave-vector $q$. A generic probe simply measures a noise $N(\omega) = \sum_q |g_q|^2 S_q (\omega)$. 

\begin{center}
\begin{figure*}[t]
\includegraphics[width = 7 in]{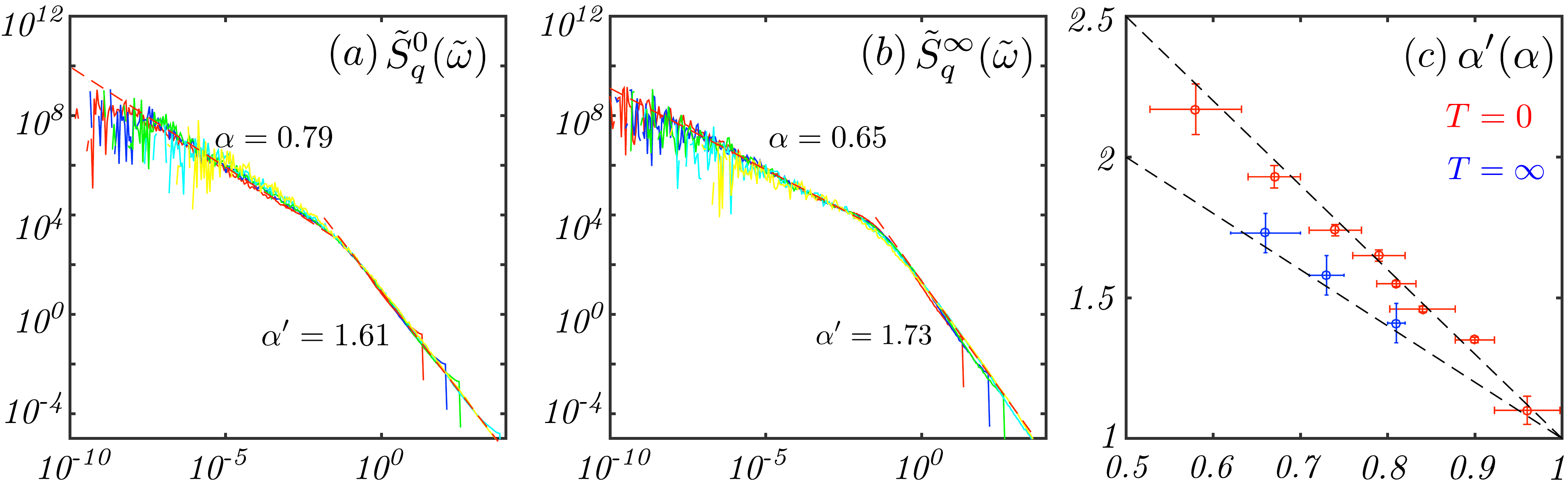}
\caption{Scaled dynamical structure factor (a) $\tilde{S}^0_q (\tilde{\omega}) = x^{1/2-1/\beta} S^0_{q}(\omega x^{1/\beta})$ at $T = 0$, (b) $\tilde{S}^\infty_q (\tilde{\omega}) = x^{-1/\beta} S^\infty_{q}(\omega x^{1/\beta})$ at $T = \infty$ for $q \in 2 \pi / [ 40 (\text{yellow}), 80 (\text{cyan}), 120 (\text{green}), 400 (\text{blue}),800 (\text{red})]$, and $x = (2\pi /40)/q$; the results shown correspond to a system of an equal proportion of F/AF bonds, and disorder strength $D = 3$; (c) scaling relations at $T = 0$ and $T = \infty$ between the power laws $\alpha$ and $\alpha'$ determining the high- and low-frequency dependence of $S_q (\omega)$ as given in Eq.~(\ref{eq:sqom}). $\alpha$ and $\alpha'$ were plotted for $D = 3$ and $\eta_i =  [1,0.75,0.5,0.25,0,-0.25,-0.5,-0.75]$ for $T = 0$, while, at $T = \infty$, the results are independent of $\eta_i$, and the data-points correspond to $D = [1,3,9]$. Decreasing $D$ and decreasing $\eta_i$ correspond to smaller values of $\alpha$. At $T = \infty$, $\alpha' \approx 1 + 2 (1 - \alpha)$, while at $T = 0$, $\alpha' \approx 1 + 3 (1 - \alpha)$; these relations are plotted as dashed lines. The optimal scaling collapse in (a) (at $T = 0$) and (b) (at $T= \infty$) was found for $\beta = 0.40$ and $\beta = 0.36$ respectively. These values are in accordance with analytically determined relation to the low-frequency noise exponent $\alpha$, i.e., $\beta = 2( 1- \alpha)$ at $T = 0$ and $\beta = 1- \alpha$ at $T = \infty$.}
\label{fig:figmain}
\end{figure*}
\end{center}

\subsection{Convergence}

It was pointed out in Ref.~\cite{westerberg} that the uniform distribution flows to a universal fixed point (at zero temperature) with a final bias $\eta_f \approx 0.26$ irrespective of its initial bias $\eta_i$, while more singular distributions, such as the ones we primarily consider do not flow to that particular fixed point. While we recover this result for initially uniform distributions in our simulations, we note that for the $1/|J|$ distributions we consider, the system also flows to fixed points (with a stable final bias $\eta_f \in(0,0.3)$, see Appendix~\ref{sec:tzeroinf}) but one that varies slightly depending on the value of the initial bias $\eta_i$, disorder strength $D$, and temperature. Regardless of the precise values of the initial bias, the form of the distribution, or the temperature, we find that the effective gaps (and the corresponding effective couplings, as well) flow to a power law distribution $P_\Delta (\Delta) \sim 1/\Delta^\gamma$, $\gamma < 1$. This validates the RG procedure because a singular power law distribution results in a \emph{typical} value of the gap $\Delta$ that is significantly smaller (by a factor $e^{1/(1-\gamma)}$) than the maximum gap $\Delta_0$ and guarantees a separation in energy scales between nearby regions. We also note that while the RG protocol results in larger and larger spin sizes over the course of the RG, these spins also couple more weakly to their neighbors (see Appendix~\ref{sec:jscorr}) and this ensures that the RG can converge to a fixed point with stable scaling properties. 

The final bias $\eta_f$ and $\gamma$ display the following trends: i) at $T =0$, for a uniform distribution, we find in confirmation with Ref.~\cite{westerberg}, that $\eta_f$ and $\gamma$ are independent of the initial bias; ii) at $T= 0$, for $1/|J|$ initial distribution, these quantities depend on the initial bias. In particular, $\gamma$ goes from $1$ to $0$ as the initial bias is varied from a purely AF to purely F chain, signaling the breakdown of the RG procedure for the purely F chain; and iii) at $T = \infty$, for both forms of distributions, $\eta_f \approx 0$ always, and $\gamma$ is independent of the initial bias but depends weakly on disorder strength $D$ for the $1/|J|$ initial distribution. Further details on these observations can be found in Appendix~\ref{sec:numchecks}. 

[Additional validity checks : i) simulations for classical spin chains using classical RSRG rules at $T = \infty$ were found to compare well with the quantum $T = \infty$ RSRG results, see Appendix~\ref{sec:classicalRG} ; ii) the structure factor obtained from the classical RG results was found to be in good agreement with that obtained from direct integration of Bloch-dynamics of the classical spins, see Appendix~\ref{sec:rgbloch}]. 

\subsection{Form of the structure factor}

A most interesting facet of the simulations is the emergence of a finite, anomalous dynamical exponent $z = 1/\beta \neq 1$,$2$. This dynamical exponent dictates that the size of clusters, $n$, and the time-scale of their dynamics, $\omega^{-1}$, scale as $n \sim \omega^{-\beta}$ [as illustrated in Fig.~\ref{fig:figrules} (c)]. Furthermore, this scaling gives rise to a piece-wise power-law behavior of the frequency dependence of the structure factor $S_q (\omega)$ for $\omega \gg q^{1/\beta}$ and $\omega \ll q^{1/\beta}$ corresponding to whether the probe's period of oscillations $2\pi/q$ is much greater than the cluster size or much smaller. Moreover, the exponent $\beta$ is directly connected to the exponent $\alpha$ in the noise spectrum $N(\omega) \sim 1/\omega^\alpha$. 

The full form of the dynamical structure factor $S_q (\omega)$, defined as $S^0_q (\omega)$ at $T = 0$ and $S^\infty_q (\omega)$ at $T = \infty$ can be extracted using a scaling collapse [Fig.~\ref{fig:figmain} (a), (b), see also Appendix~\ref{sec:qscalingcheck} for verification of $q-$dependent scaling] and is found to be 

\begin{align}
& S^0_q(\omega) = q^{1/2-1/\beta}  g_0 \left( \frac{\omega}{q^{1/\beta}} \right) = \begin{cases} 1/\omega^{1-\beta/2} & \omega \ll q^{1/\beta}, \\ q^2/\omega^{1+ 3\beta/2} & \omega \gg q^{1/\beta}, \end{cases} \nonumber \\
& S^\infty_q(\omega) = q^{-1/\beta} g_\infty \left( \frac{\omega}{q^{1/\beta}} \right) = \begin{cases} 1/q \omega^{1-\beta} & \omega \ll q^{1/\beta}, \\ q^2/\omega^{1+ 2 \beta} & \omega \gg q^{1/\beta} ,\end{cases}
\label{eq:sqom}
\end{align}
where $g_0 (x)$ and $g_\infty (x)$ are arbitrary scaling functions whose limits $ x \ll 1$, $x \gg 1$ have been verified numerically. (The limiting forms of the structure factor are also derived using an analytical approach in Sec.~\ref{sec:scalingapproach}.) The low-(high-)frequency power-law exponent in both cases is identified as $\alpha$ ($\alpha'$).  As we had mentioned before, $S_q (\omega)$ shows a power-law frequency tail of the form $1/\omega^\alpha$. It is curious to note that, at high temperatures, the noise exponent deviates from $1$ by the inverse of the dynamical exponent, i.e., $\alpha = 1 - 1/z$ while at low temperatures, the appropriate relation is $\alpha = 1 - 1/2z$. The frequency dependence of the noise at higher frequencies (characterized by exponent $\alpha'$) is also related in a non-trivial way to the dynamical exponent that differs at low and high temperatures. 

Using the scaling form of the structure factor, one can easily find that a typical probe that measures the noise $N(\omega) = \sum_q |g_q|^2 S_q(\omega)$ will not observe any lower bound in the $1/\omega$-like behavior of the noise. In particular, for frequencies $\omega \ll q^{1/\beta}_0$, where $q_0$ characterizes the probe's resolution so that $g_q (q < q_0 ) \approx const.$, the noise spectrum will inherit the low-frequency behavior of this finite-$q$ structure factor. 

As for the precise values of these exponents, we find at $T = 0$, for the $1/|J|$ initial distribution, that the low-frequency noise-exponent $\alpha$ varies from $1$ ($\beta = 0$) for the AF chain to approximately $0.5$ ($\beta = 1$) for the nearly F chain. The AF result of $\alpha \approx 1$ can be understood from the point of view that the purely AF Heisenberg spin chain flows to an infinite-randomness fixed point with a distribution of couplings $P(J) \sim 1/J$. In the nearly AF case, these couplings also directly give the gap $\Delta = J ( 1 + | s_1 - s_2 | ) \approx J$ since the system forms only singlets and the spins don't grow in magnitude over the course of the RG. Since the noise generation happens at the frequency $\Delta$ (in particular, at every RG step, the maximum gap $\Delta_0$), it simply inherits the power law of the coupling distribution. (The connection between the noise power $\alpha$ and the power law of the gap distribution is detailed in Appendix \ref{sec:exprelate})

As we introduce more ferromagnetic bonds into the spin chain, the low-frequency noise exponent gradually approaches $0.5$, and the distribution of effective gaps becomes less singular (accompanied by a reduction in the fidelity of the RG). For an initial bias $\eta_i \approx -0.75$, the distribution of gaps is almost uniform indicating a failure of the RG. Note that, at $T = 0$, $\alpha \rightarrow 0.5$ is accompanied by a dynamical exponent $z = 1/\beta \rightarrow 1$ (recall that exponents are related at $T = 0$ via $\alpha = 1 - 1/2z$), and one can show (see Appendix~\ref{sec:roleofresonances}) that in such a case the RSRG fails due to the proliferation of faraway resonances. 

To gain more insight into the behavior of nearly F chain, we perform a Holstein-Primakoff (see Appendix~\ref{sec:hpdiffusion}) expansion on top of the fully polarized ground state of a purely F chain. Unlike Eq.~(\ref{eq:sqom}), the structure for the purely F spin chain is peaked at a finite frequency $\omega$ for wave-vector $q$ that scales as $\omega \sim q^2$ as expected for magnons. However, these magnons are heavily damped; the width of the magnon peak also scales as $q^2$. Due to the large damping, these spin waves are `localized' in the sense that they exhibit a finite inverse participation ratio [given by $\sum_i n_i^2 / (\sum_i n_i)^2$ ; $n_i$ is the spin wave density at site $i$ of the chain]. If we interpret this finite inverse participation ratio as a localization length, we find that it diverges in the low-frequency limit as $\xi (\omega) \sim 1/\omega$; this is in agreement with the observation that spin clusters scale as $n(\omega) \sim 1/\omega$ ($z = 1$) for the nearly F chain in the RG at $T = 0$. (In two dimensions, the Holstein-Primakoff analysis shows that the spin-wave peaks are sharper and delocalized; the results are in agreement with effective-medium approximation based approaches~\cite{edwardsferro,yatesferro}.) 

At infinite temperature, $\alpha$ is independent of the the initial bias, but depends weakly on the disorder $D$, approaching $1$ as disorder is increased [a crossing beyond $1$ is not observed, see Fig.~\ref{fig:figmain} (c) where $\alpha$ and $\alpha'$ are plotted]. The range of the low-frequency noise-exponent is, as in the $T = 0$ case, $0.5<\alpha<1$. In Sec.~\ref{sec:generalizedapproachandmbl} we will argue that the infinite-temperature results we obtain are natural for any system exhibiting anomalous diffusion. Moreover, we will see that the special case of $\alpha = 1$ (or $1/\omega$ noise, precisely) is singular enough that it leads to the absence of relaxation of non-conserved finite-$q$ modes signaling many-body localization. It seems reasonable, in the light of these arguments, that the exponent $\alpha$ should approach $1$ as the disorder strength $D$ is increased [as we find, see Fig.~\ref{fig:figmain} (c)]. 

We now recapitulate the findings for the \emph{strongly} disordered Heisenberg spin chain with initial coupling distribution $P_0(\abs{J}) \sim 1/\abs{J}$ ,$\abs{J} \in [ 1, e^D ]$.  At $T = 0$, we find that the system has three distinct regimes according to the initial bias: a) the purely AF spin chain which flows towards infinite-randomness and has a divergent dynamical exponent and $1/\omega$ noise spectrum; b) the purely F spin chain which has peaked spectral functions associated with heavily damped magnons whose size $n(\omega)$ diverges as $1/\omega$; and c) a wide range of mixed AF/F spin chains which flow towards non-universal strong-randomness fixed points (thus, accessible by RSRG) and whose structure factor is of the form in Eq.~(\ref{eq:sqom}). These chains exhibit $1/\omega^\alpha$ noise for $0.5 < \alpha < 1$, which arises due to spin clusters whose size $n(\omega)$ diverges as $1/\omega^{1/z}$ with $z > 1$; this divergence is \emph{slower} than $1/\omega$ found for the purely F chain. At $T = \infty$, the physics of the spin chain is independent of the proportion of AF/F bonds. It exhibits a $1/\omega^\alpha$ noise spectrum, again with $0.5 < \alpha < 1$ where $\alpha$ increases as $D$ is increased. Note that at $T = \infty$, the dynamical exponent is related to $\alpha$ via the relation $\alpha = 1 - 1/z$, implying that $z  \ge 2$ at $T = \infty$. 

We can qualitatively glean the form of the structure factor at finite temperatures by considering a straightforward extension of the RSRG protocol: we follow the zero-temperature RSRG rules for eliminating modes with frequencies $\omega \gtrsim T$ (since we expect these modes to be populated primarily in their ground state configurations), and infinite-temperature rules for eliminating modes at frequencies $\omega \lesssim T$ (since we expect these modes to be sufficiently excited). As we discuss below, such a protocol leads to the conclusion that the dynamics of the system resemble the $T = \infty$ ($T = 0$) behavior for frequencies $\omega \ll T / 2^{z_0}$ ($ \omega \gg T / 2^{z_0}$ ), where $z_0$ is the dynamical exponent of system (with a given initial bias $\eta_i$) found at $T = 0$. Over the course of eliminating high-frequency modes ($\omega \gtrsim T$), the gap distribution develops a power-law form with an exponent that is expected for the zero-temperature system. The zero-temperature RG protocol continues until $\Delta_0 \approx T$. Beyond this scale, the RG continues with infinite-temperature rules, which begin to have an effect on the gap distribution only when a significant fraction, say $1/2$ of the spins have been eliminated. Using the dynamical exponent $z_0$, the maximum gap $\Delta_0$ at this length scale is given by $\Delta_0 = T/2^{z_0}$. Thus, we expect that the infinite-temperature form of the structure factor appears only for frequencies below $\omega \lesssim T/2^{z_0}$. 

Our findings are summarized in the phase diagram of Fig.~\ref{fig:figalphaz}. 

\section{Scaling approach to Structure factor}
\label{sec:scalingapproach} 
\subsection{Scaling distribution $P_{F (AF)}$}

We now explain the results in Eq.~(\ref{eq:sqom}) by extending the arguments of Westerberg et al. (Ref.~\cite{westerberg}) to evaluate the dynamical structure factor. They find that the RG flow eventually converges to a fixed point where the F/AF bias stabilizes and the probability distribution for F (AF) bonds is given by the scaling form $P_{F (AF)} (\Delta, S_L , S_R) \sim \frac{1}{\Delta_0^{1 - \beta}} Q_{F(AF)} ( \Delta \Delta^{-1}_0, S_L \Delta^{\beta/2}_0, S_R \Delta^{\beta/2}_0 )$, where $\Delta_0$ corresponds to the maximum gap at any step of the RG, while $S_L$, $S_R$ are the left and right spins across the bond of frequency $\Delta$. Note that, at $T = 0$, $\Delta$ is defined as the energy difference between first excited state from the ground state of the F/AF spin pair, while, at infinite temperature, the notion of a gap still holds---entropic considerations dominate and the relevant gap is approximately the gap associated with the largest few total-spin states. The form implicitly assumes a dynamical exponent $z = 1/\beta$; the scaling of the spin size with the exponent $-\beta/2$ occurs because it scales as the square-root of the cluster size $n$ (as long as the chain is not purely F or AF); $n$ scales with the exponent $-\beta$) at both high and low temperatures. The reason that both F and AF spins scale in the same way is attributed to the fact that proportion of F and AF bonds in the limit that the RG converges is a finite, intermediate value. Additional requirements of finiteness of $\avg{\Delta}/\Delta_0$ and normalization fix the form of the distribution. 

The emergence of a finite dynamical exponent is explained as follows: in $k$ RG levels, $\sim 2^k$ microscopic spins are combined into a single cluster, while the energy gap, expected to reduce by a factor $r$ at each level, reduces from $\Delta_0$ to $\Delta_0/r^k$; the dynamical exponent $z$ is then readily found to be $z =  \ln{r}/\ln{2}$. Note that even though our simulations with more singular initial bond-distributions exhibit non-universal values of the exponent $z = 1/\beta$, the convergence of the RG towards a fixed value of the final bias indicates that the notion of a scaling distribution as discussed above still applies.  

\subsection{Calculation of $S_q (\omega)$}
\label{sec:scalc}
From the  scaling forms of the distribution functions $P_{F(AF)}$, the dynamical structure factor $S_q (\omega)$ is given by 

\begin{align}
S_q (\omega) =& \sum_{a = F,AF} \int  d\Delta_0 dS_L dS_R \delta(\omega - \Delta_0) N (\Delta_0) \nonumber \\
&  \times M(S_L,S_R)  F(qn_0) P_a (\Delta = \Delta_0, S_L, S_R) 
\label{eq:sqomtheory}
\end{align}

The above integral simply reflects the discussion of the noise calculation procedure: at every step of the RG, a local integral of motion combing spins $S_L$ and $S_R$, with a gap $\Delta = \Delta_0$ is eliminated, producing noise of magnitude $M(S_L,S_R) F(q n_0)$ at frequency $\omega = \Delta_0$. Here, as mentioned before, $M(S_L,S_R)$ corresponds to matrix elements of the spin operators associated with the transition at frequency $\omega = \Delta_0$, while $F(qn_0)$ modifies this amplitude depending on how the probe couples to the cluster generating the noise; this depends on the ratio of the size $n_0 (\Delta_0)$ of the typical clusters at maximum gap $\Delta_0$ and the probe wave-length $2\pi/q$. $N(\Delta_0) \sim 1/n_0 \sim \Delta_0^{\beta}$ denotes the number of bonds that remain at the cut-off scale $\Delta_0$. 

We now use Eq.~(\ref{eq:sqomtheory}) to evaluate the form of the structure factor and show that it reproduces the results in Eq.~(\ref{eq:sqom}). 
The noise magnitude $M(S_L,S_R) \sim | \vec{S}_-|^2$ and, consequently, at any finite temperature (as any generic combination of spins $S_L$ and $S_R$) scales as the square of the typical spin size, i.e., $M \sim s_0^2 \sim \Delta_0^{- \beta}$. However, at zero temperature, the scaling changes to $M \sim s_0 \sim \Delta_0^{-\beta/2}$ (see Appendix~\ref{sec:noisecalc}). 

Next we consider the scaling of the form factor $F(q n_0)$. To arrive at the results in Eq. (\ref{eq:sqom}) starting from Eq.~(\ref{eq:sqomtheory}), we need to show that at both $T =0 $ and $T = \infty$, and frequencies $\omega \gg q^{1/\beta}$, $F(qn_0) \sim q^2 n_0^2$, while for frequencies $\omega \ll q^{1/\beta}$, at $T  = 0$, $F(qn_0) \sim \text{const.}$ and at $T = \infty$, $F(q,n_0) \sim 1/q n_0$. Let us note that the distinction between the two regimes comes from the fact that at high-frequencies, the clusters are smaller than the typical length $1/q$ of the modulations of the probe, while at lower-frequencies, the clusters are larger in comparison. So, the noise generated by two clusters $A$ and $B$ much smaller than the probe scale $1/q$ comes with an form-factor $F(qn_0)$ that scales as the square of the gradient $q$ of the probe, that is, $F(q n_0) \sim q^2 n^2_0$. (Recall that the form factor is the square of the difference of the couplings of these individual clusters $A$ and $B$ to the external probe.)

We now consider the scaling of $F(q n_0)$ at low frequencies ($\omega \ll q^{1/\beta}$) where clusters are much larger than the probe scale $1/q$. At low temperatures, clusters are composed of a large number of singlets, and consequently, cluster sizes and the bond lengths scale in the same way (as $\sim n_0$, see Ref.~\cite{westerberg}). When two such clusters are merged, the relative phases of the probe coupling are effectively scrambled. Thus, in this regime, at low temperatures, $F(qn_0)$ is $q$-independent.

At $T = \infty$, we find the scaling of $F(qn_0) \sim 1/qn_0$ using the following two facts---i) singlets are entropically unlikely at high temperatures; clusters are compact (nearly contiguous arrays of spins that are not part of singlets) and bond-lengths are of the order of the microscopic scale and; ii) spin-clusters point in arbitrary directions; as a result, one can show that variance $\sigma(g_+)$ of the probe coupling $g_+$ of a cluster composed of two smaller clusters with couplings $g_A$ and $g_B$ respectively is precisely equal to the probe form factor associated with combining these clusters (see Appendix~\ref{sec:noisefactinf}). The first of these implies that the mean of the probe coupling $g_A$ of cluster $A$ of size $n_0$ can be approximated by $ \abs{\avg{g_A}} \sim \int^n_0 \; dx \; \cos (q x) / n \sim 1/qn_0$ for $q n_0 \gg 1$. The second result shows that the scaling of the probe form factor is given by the scaling of the variance of the probe couplings; $F(q n_0) \sim \sigma(g_+) = \avg{ (g_A - g_B )^2/4} \approx [ \sigma(g_A) + \sigma(g_B) ] /4$, where, to obtain the last (approximate) result, we assumed that the squared-mean $\avg{g_A}^2 \sim 1/(qn_0)^2$ is negligible in comparison to the variance $\sigma(g_A)$ in the limit $qn_0 \gg 1$. If we reasonably assume that clusters $A$ and $B$ of similar size $\sim n_0$ (and consequently, similar variance) are combined typically, then the this immediately leads to the result $F(q n_0) \sim \sigma (g_+) \sim 1/qn_0$ (which also justifies neglecting the mean values $\avg{g_A}$,$\avg{g_B}$) as mentioned above. 

With the aid of the specific form of the distribution function, the noise amplitude and the form factors discussed above, we can perform the integration in Eq.~(\ref{eq:sqomtheory}) by first homogenizing all flow parameters in favor of the maximum gap $\Delta_0$, which can then be integrated directly to yield the $q$,$\omega$ dependent behavior. This yields the results in Eq.~(\ref{eq:sqom}).

\section{Generalized Diffusion and the Structure Factor} 
\label{sec:generalizedapproachandmbl} 
 The derivation of the structure factor presented in the previous section relies on specific details of the model. We now present a general ansatz for the spin-diffusion propagator in a system exhibiting anomalous diffusion and show that it reproduces the form of the structure factor in Eq.~(\ref{eq:sqom}) at high temperatures. (See also Ref.~\cite{agarwal2014anomalous}.)
 
\subsection{Ansatz for the anomalous diffusion propagator}

The spin propagator describes the decay of the spin density $S(x,t)$ at any point $x$ and time $t > t'$ given the spin density profile at all points $x'$ at some fixed $t'$, that is, $S(x,t) = \int dx' G(x-x', t-t') S(x', t')$, where, by definition, $G(x-x', 0) = \delta(x-x')$. Note that we do not keep the tensor structure of the Green's function because off-diagonal components vanish by symmetry and there is no preferred ordering direction in one-dimension for any non-zero concentration of AF bonds at any temperature. (In higher dimensions, the approach is directly applicable above any ordering temperature.) In the case of regular diffusion, the Green's function satisfies the diffusion equation, and, in the Laplace domain, is given by $G_q(\omega) = 1/(-i\omega + Dq^2)$. This diffusive propagator has two salient features: i) $\lim_{\omega \rightarrow 0} (- i \omega) G_{q=0}(\omega) = 1$ implying that the $q = 0$ mode does not relax (that is, total spin/density is conserved), and ii) it has a pole at finite imaginary frequencies for all finite-$q$ modes, which as a consequence, relax exponentially in time.  

We now generalize this propagator to the case of anomalous diffusion, where the system exhibits an anomalous scaling (the dynamical exponent $z  = 1/\beta \neq 1,2$) between $q$,$\omega$: $q \sim \omega^\beta$ -
\be
G_q(\omega) = 1/[-i\omega + q^{1/\beta} f(\omega/q^{1/\beta})],
\label{eq:anom}
\ee
where $f$ is some well-behaved function which ensures conservation of the total spin [$G_{q =0}(\omega) \sim i/ \omega$]. 

\subsection{Calculation of the structure factor at $T = \infty$}

We would like to use the Green's function postulated above to calculate the structure factor $S_q (\omega)$. First, we introduce the dynamical susceptibility $\chi_q (\omega)$ which is the usual Kubo-response of the spin-density $\vec{S}_q = \int dx \vec{S} (x) e^{i q x}$ to a field $h_q$ that couples to $\vec{S}_{-q}$. That is, $\chi_q (\omega)$ is the Laplace-transform of $\chi_q (t-t') = - i \theta (t-t') \avg{\left[\vec{S}_q(t).,\vec{S}_{-q}(t') \right]_-}$. The dynamical susceptibility is connected to the propagator $G_q (\omega)$ via the relation $\chi_q (\omega) = \chi^0_q [ 1 + i \omega G_q(\omega) ]$, where $\chi^0_q$ is the static susceptibility at wave-vector $q$. The result can be rationalized as follows: the measurement of the dynamical susceptibility is carried out by slowly ramping up the perturbation $h_q (t) = h^0_q e^{\epsilon t}$ (coupled to $S_{-q}$) for times $t < 0$ and observing the relaxation of the spin-density $S_q$ for subsequent times. One can think of such an experiment as one that sets up a spin-density $\chi^0_q h^0_q$ by time $t = 0$, and whose subsequent relaxation is given by the diffusion propagator, i.e., $\avg{S_q (\omega)} = \chi^0_q h_q G_q (\omega)$ (in Laplace-domain; note that $\avg{S_q (\omega)}$ denotes the expectation value of the spin-density operator and not the structure factor). Alternatively, we can appeal to the definition of the dynamical susceptibility to directly find $\avg{S_q (\omega)} = [\chi_q (\omega) - \chi^0_q] h^0_q / i\omega$ (see Sec.~(7.14) of Ref.~\cite{chaikin2000principles}). These alternative interpretations yield the relation between $G_q (\omega)$ and $\chi_q (\omega)$. Finally, the structure factor is related to the imaginary part of the dynamical susceptibility using the (fluctuation-dissipation) relation $S_q (\omega) = \coth{\left(\omega/2T\right)} \text{Im} \left[ \chi_q(\omega) \right]$. 

At high temperatures ($\omega \ll T$), the Green's function in Eq.~(\ref{eq:anom}) yields a structure factor $S_q (\omega) = T \chi^0_q q^{-1/\beta} f(\omega/q^{1/\beta}) / [ (\omega/q^{1/\beta})^2 + f(\omega/q^{1/\beta})^2]$. It is easy to confirm that the above form can be represented as $S_q (\omega) = const. \times q^{-1/\beta} g_\infty (\omega/q^{1/\beta})$, in agreement with the result for $S^\infty_q (\omega)$ in Eq.~(\ref{eq:sqom}) if we assume that $T \chi^0_q$ has a finite limit for small $q \ll 1$. This is expected for any physical system with short range interactions, at high temperatures. 

To complete the argument, we now determine the limiting forms of the scaling function $g_{\infty}(y = \omega/q^{1/\beta})$, that is, its high (low)-frequency limits $y \gg 1$ ($y \ll 1$). First, note that the (real part of the) optical conductivity $\sigma_q (\omega) = \avg{ \left[ J_q (\omega) , J_{-q} (-\omega) \right] } / i \omega $ can be computed from the structure factor using the relation $\text{Re} \left[ \sigma_q (\omega) \right] = \frac{\omega}{q^2} \tanh{\left(\omega/2T\right)} S_q (\omega)$; the result (at any non-zero frequency) follows from the continuity relation $ i\omega J_q (\omega) - i q S_q (\omega) = 0$ and using the fluctuation-dissipation relation discussed above. Straightforward calculation using these results yields $\sigma(q = 0,\omega) = c(q=0) \omega^{1 - 2 \beta}$, where the pre-factor $c(q)$ is finite and non-zero only if $g_{\infty} (y) \sim y^{-1 - 2 \beta}$ for $y \gg 1$. Since there is no reason for the optical conductivity to vanish or diverge at finite frequencies in a high-temperature system, we expect this condition yields the scaling of the function $g_\infty (y)$ in the high-frequency limit. The argument for the low-frequency limit is as follows. The form of the Green's function ensures $G(x = 0, t) \sim 1/t^{\beta}$. On this basis, we expect large $q$ modes should decay algebraically as $\sim 1/t^\beta$ as well. This implies that $f(y \ll 1) \sim y^{1 - \beta}$, and consequently, $g_\infty(y \ll 1) \sim y^{-1 + \beta}$. These results together reproduce the complete form and limits of the structure factor $S^\infty_q (\omega)$ ($T = \infty$) in Eq.~(\ref{eq:sqom}). Because of the generality of this derivation, we expect that any system exhibiting anomalous diffusion must have a structure factor that scales as described in Eq.~(\ref{eq:sqom}). 

\subsection{Failure of linear response at $T  =0$}

The generalized-diffusion ansatz fails to explain the $T = 0$ structure factor. In particular, the low-frequency power-law $S_q(\omega)\sim 1/\omega^\alpha$, with $0\lt \alpha \lt 1$ is impossible to obtain with this Green's function form. Furthermore, according to the thermodynamic sum rule,  $\chi(q) = \int d\omega \chi''(\omega)/\omega$ ($\chi''(q,\omega) = \text{sign}(\omega) S_q (\omega)$), the static susceptibility is expected to diverge even at finite $q$ making the derivation suspect. Hence, at $T = 0$, we are forced to conclude that either linear response does not work, or the scaling hypothesis for the Green's function breaks down, or both. In what follows, we provide a heuristic argument for the breakdown of linear response by showing the the divergence of the static susceptibility at finite-$q$. 

Following Ref.~\cite{westerberg}, we imagine applying a perturbing field $h_q e^{i q x}$ to the chain. This perturbation cuts off the RG flow at a scale where the energy gap $\Delta_0$ is comparable to the Zeeman energy of typical, polarizable (of size $n \lesssim 1/q$) spin-clusters; that is, the RG is cut-off at $\Delta_0 \sim h_q \; \text{min} \left[\sqrt{n_0},1/\sqrt{q} \right]$, where we note that the typical cluster size at an energy scale $\Delta_0$ is given by $n_0 \sim \Delta_0^{-\beta}$, and possesses a spin $\sim \sqrt{n_0}$. Crucially, the magnitude $h_q$ determines the point at which the RG is cut-off (in particular, whether $n_0 > 1/q$ or $n_0 < 1/q$ at this scale) and consequently, the response of the system. Let us define $n_0 (h_q)$ and $\Delta_0 (h_q)$ as the amplitude $h_q$-dependent length and energy scales at which the RG is cut-off. The condition $n_0 (h_q) \lessgtr 1/q$ can be alternatively cast (using the scaling results) as $q^{1/2 + 1/\beta} \lessgtr h_q$, and both limits of this condition may be experimentally relevant. 

 We first analyze the case $q \gg 1/n_0 (h_q), \Delta_0 (h_q) \sim h_q /\sqrt{q}$ first. We note that the magnetization $m$ of polarized clusters of size $n$ is $\sqrt{n}$ for $n < 1/q$ and small otherwise. If we assume that the distribution $D(n)$ of the cluster size $n$ satisfies $D(n < n_0) \sim 1/n_0$ and $D(n>n_0)$ is negligible (the form of $D(n)$ is numerically justified, see Appendix~\ref{sec:dnform}), then the average magnetization is given by ${m_q} = n^{-2}_0 (h_q) q^{3/2} = q^{-3/2 -\beta} h_q^{2 \beta}$. Thus, for $2 \beta < 1$, the static susceptibility $\chi^0_q = {m_q}/h_q |_{h\rightarrow 0}$ diverges even at finite $q$.  
 
 For the opposite limit of $q \ll1/n_0 (h_q), \Delta_0 (h_q) \sim h_q \sqrt{n_0} $, the magnetization is simply $m_q = m_{q = 0} \sim 1/\sqrt{n_0}$ and the susceptibility diverges for all values as $\beta$ as, $\chi^0_q = {m_q}/h_q |_{h\rightarrow 0} = h^{-2 / (2 + \beta)}$, in agreement with the $q = 0$ result of Ref.~\cite{westerberg}. 
 
 Thus, the static susceptibility generically diverges at $T = 0$ in this system, although the precise nature of the divergence is controlled by the condition $q^{1/2 + 1/\beta} \lessgtr h_q$. Let us note that, a divergent static susceptibility at certain wave-vectors for clean F (at $q=0$) or AF (at $q = \pi$) systems is not surprising---a small perturbation with the correct wave-vector results in a macroscopic reduction in free-energy and consequently, a divergent response. Since, in our disorder system, the probability distribution of cluster sizes, $D(n)$, has significant weight for all clusters of size $n < n_0$ (and $n_0$ diverges in the zero-frequency limit), our system has a thermodynamically-relevant presence of approximately independent spin clusters of all sizes. This results in a divergent response to perturbations over a broad range of wave-vectors. 

\subsection{Relation of $1/f$ noise and many-body-localization at finite temperatures} 

 For frequencies $\omega \ll T / 2^{z_0}$ (where $z_0$ is the dynamical exponent found for the system at $T = 0$, and is non-finite only in the purely AF case), we expect the behavior of the structure factor to be given by the infinite-temperature form of the structure factor we compute. This structure factor has a noise spectrum $1/f^\alpha$, where $\alpha \le 1$. We can now reverse the arguments of the previous section to conclude that the spin-propagator $G( q, \omega) \sim 1/\omega^{\alpha}$ for frequencies $\omega \ll T/2^{z_0}$,$q^{1/\beta}$ where $\beta = 1 - \alpha$. Fourier transforming this result implies directly that $G(q,t) \sim 1/t^{1-\alpha}$, that is, the spin-density at wave-vector $q$ decays algebraically with a power $1 - \alpha \ge 0$, where, in particular, if $\alpha = 1$, the mode does not decay completely even at infinite time. The inability of certain non-conserved quantities to decay to equilibrium values is a defining characteristic of many-body-localization, and it is reasonable to posit, on the basis of this discussion, that when the noise power law is precisely $\alpha = 1$ (or, noise is $1/\omega$), the system is at the verge of becoming many-body-localized. 
 
 A complimentary picture is obtained by looking at the dynamical exponent $1/\beta = 1/ (1-\alpha)$ (at finite $T$). In the course of RG, when a cluster joins another cluster to form a supercluster, its spin begins to precess around the direction of the total spin. Further mergers into ever larger clusters add new axes of precession with ever decreasing frequencies, which leads to scrambling of the spin polarization. The RG suggests that, for the disordered Heisenberg system, the time scale for the phase scrambling over a length scale $l$ scales as a power law $t_l \sim l^{1/\beta}$. This implies that for $\alpha < 1$, it takes only polynomial time in the distance to transport spin across this distance, while, for $\alpha = 1$, it becomes exponential (or worse) in the distance, which is suggestive of MBL. 
 
 Since we do not observe such a case (with $\alpha =1$) in our numerical simulations, we conclude that the disordered Heisenberg spin chain is not localized. Nevertheless, we note that as the disorder strength is increased, the noise exponent $\alpha$ tends towards $1$. Moreover, at low temperatures, the power law $\alpha = 1$ is obtained for the purely AF chain which flows towards the \emph{infinite}-randomness point---here the integrals of motion are exact since the RSRG is precise in the low-energy limit.
   
\section{Simulations in two-dimensions}
\label{sec:2dsimulsmain}

Simulations in two-dimensions are performed analogously to the one-dimensional case except that the number of neighbors of an effective spins can change over the course of the RG. Maintaing this extended network introduces a computational cost that restricts simulations to smaller system-sizes in two dimensions; we consider square systems of spins on a $50 \times 50$ lattice and `strips' of spins on a $500 \times 6$ lattice. The growth in the connectivity of the network over RG steps also reduces the efficacy of the RG because achieving a separation of scales becomes less probable. In particular, at $T = 0$, when singlet-formation is preferred, the clusters become highly non-compact, extensive objects with a large number of neighbors, and consequently the fidelity, which is the ratio of maximum energy scale in the system compared with the energy scales amongst the spins neighboring the maximally-coupled spins, does not rise substantially over the course of the RG procedure, even if the distribution of couplings assumes a power law form (see Appendix~\ref{sec:2dsimul}). At high-temperatures, the RG is more robust since singlets are entropically unfavorable, and clusters grow uniformly keeping the number of neighbors relatively unchanged over the course of the RG. In what follows, we restrict ourselves to a discussion of the $T = \infty$ simulations on the two-dimensional square and strip geometries. 

For a spin network arranged in a square geometry, we find that the structure factor exhibits a scaling collapse in accordance with the generalized-diffusion form [$S^\infty_q (\omega)$ in Eq.~(\ref{eq:sqom})] with a dynamical exponent $z \approx 2$; this is accompanied by a low-frequency $1/\sqrt{\omega}$ tail in the finite-$q$ structure factor (see Appendix~\ref{sec:2dsimul}). We note that the dynamical exponent $z =2$ is also associated with regular diffusion; the important difference here is that the diffusive structure factor is flat at low frequencies. Since the fidelity does not improve significantly over the course of the RG for systems with square geometries, we cannot conclude definitively on the nature of dynamics of such systems. 

From the point-of-view of experiments measuring flux noise in SQUID systems, we consider the case of a two-dimensional strip of spins (the width of the conducting strip of the SQUID circuit is typically much smaller than the length of the SQUID loop). In this case, the RSRG finds that the structure factor possesses a $1/\sqrt{\omega}$ power-law frequency-dependent behavior which eventually, at lower frequencies, gives way to an anomalous power-law behavior, $1/\omega^\alpha$; $\alpha \approx 0.70$ (see Fig.~\ref{fig:2dstrip}). The reason for this is self-evident; the clusters contributing below a certain frequency are wider than the width of the spin network and can be thought to be lined up in an effectively one-dimensional array; thus, the inheritance of the anomalous power-law behavior from one-dimension is unsurprising. 

\section{Summary and Outlook} 
\label{sec:summary}

In this work, we extended the RSRG protocol with rules for the renormalization of the couplings of a generic probe of spin fluctuations. This allowed us to compute noise spectrum measured by such a probe, and in particular, the dynamical structure factor $S_q (\omega)$. We find, for one- and quasi-one-dimensional systems with exchange-coupling magnitudes distributed according to a $1/|J|$ distribution, and a wide range of initial composition (proportion) of F/AF bonds, that the RG converges at both $T = \infty$ and $ T = 0$ to non-universal strong-randomness fixed points associated with an anomalous dynamical exponent $z$. This dynamical exponent is associated with the spectral characteristics of the noise power $\alpha$ at low frequencies: $\alpha = 1 - 1/z$ at $T = \infty$ and $\alpha = 1-1/2z$ at $T = 0$. We also find a non-trivial scaling of the high frequency end of the structure factor [see Eq.~(\ref{eq:sqom})] and relate the power law of the frequency dependence in this regime to the low frequency noise power $\alpha$. 

These findings were justified for both high and low temperatures by postulating a scaling form of the probability distribution function $P_{F(AF)}$ governing the distribution of F(AF) bonds with a given coupling, and at energy scales where the RG has converged. Additionally, we showed how an ansatz that generalizes the spin-diffusion propagator, reproduces the complete form of the structure factor at high temperatures, from which we conclude that the results obtained (including the presence of $1/\omega^\alpha$ noise) are universal and extend generally to systems exhibiting anomalous diffusion. We explain that the failure of this approach in the zero-temperature case occurs due to the failure of linear-response theory for the random-bond Heisenberg spin chain at this temperature: even at finite wave-vectors, static susceptibility of the chain diverges. 

From the point of view of interest in many-body-localization physics, our work raises an important question---is the phase we find non-localized but also non-ergodic? (Such a phase was proposed in Ref.~\cite{ioffe_nonergodic}.) The RSRG protocol creates a macroscopic number integrals of motion (albeit increasingly non-local integrals) which implies the breakdown of ergodicity. However, since the RSRG protocol suggests a flow towards a strong-randomness fixed point, and not \emph{infinite}-randomness, the RG protocol is not exact and neither are the integrals of motion it creates. This does not exclude the possibility that integrals of motion do exist, and one needs to devise a more elaborate form of the RSRG protocol to correctly capture them. Thus, the status of ergodicity in our model is unclear. 

Irrespective of the resolution of this discussion, let us note that the system is certainly not in a many-body-localized phase. As was mentioned above, the $1/\omega^\alpha$ ($\alpha < 1$) form of the finite-$q$ structure factor implies that all spin-waves decay in time as $1/t^{1-\alpha}$. As long as $\alpha < 1$ (strictly), we expect polarization in the system to relax (to equilibrium). A complimentary picture is obtained by looking at the dynamical exponent $1/\beta = 1/ (1-\alpha)$ (at finite $T$) which governs the scaling between the time $t$ it takes for spin phase to get scrambled over a distance $l$; $t \sim l^{1/\beta}$. As $\alpha \rightarrow 1$, the scaling changes from polynomial to exponential (or worse) in distance, which again suggests that the localization transition occurs at $\alpha = 1$. Since we never observe this value for $\alpha$ in the infinite-temperature simulations (although, increasing disorder strength $D$ is seen to slowly drive the exponent $\alpha$ towards 1), we conclude that the disordered Heisenberg spin chain is not localized at finite temperatures. 

Finally, we note that the noise generated by these random-bond Heisenberg spin chains with a $1/|J|$ distribution captures some interesting facets (as seen in Fig.~\ref{fig:figmain} for the experimentally relevant case of equally proportioned F/AF system, see also Appendix~\ref{sec:tzeroinf}) of experiments observing flux noise in SQUIDs: i) the noise-exponent decreases with increasing temperature; ii) the magnitude of the noise is fairly temperature independent over a large frequency range; and iii) the point of constancy in the magnitude of noise spectrum (as a function of temperature) is about 9 orders of magnitude smaller than the microscopic energy scales. The caveat is, of course, that the network of spins in SQUIDs is two-dimensional and these simulations were performed for spin chains---due to computational costs, we cannot perform simulations on equally large two-dimensional systems and are, consequently, limited in the frequency range of the calculated noise. Nevertheless, the simulations performed on elongated two-dimensional strips of spins show, below a certain frequency, an anomalous power-law frequency-dependence akin to that seen in one-dimensional chains. We believe that this may be a reasonable representation of the experimental setup.   

In future work, we aim to extend our RSRG protocol to systems with anisotropic dipolar couplings and (or) random local magnetic fields that create an additional preferred axis amongst the spins; such couplings are known to change the dynamics of system significantly~\cite{Spinglassbook,Shnirman}, making it resemble the dynamics of disordered Ising spins, which furthermore, are likely to be more amenable to a RSRG analysis even in higher dimensions. Moreover, in one-dimension, such models are known to exhibit the many-body localization transition~\cite{yaodipolarmbl,palmbl2010} and will thus allow us to further probe the connection of $1/\omega$ noise with the many-body-localization transition. It would also be interesting to examine the presence or absence of $1/\omega^\alpha$ noise in other systems where anomalous dynamical exponents are observed; such as non-Fermi liquids~\cite{senthilnonfermi,varma2002singular} and Luttinger liquids~\cite{giamarchi1d}. 

\section{Acknowledgements}

The authors acknowledge useful discussions with E. Altman, S. Gopalakrishnan, D. Huse, M. Knap, S. Kos, M. Lukin, V. Oganesyan, D. Pekker, A. Rahmani, G. Refael and N. Yao. K.A. and E.D. support from NSF grant DMR-1308435, Harvard-MIT CUA, the ARO-MURI on Atomtronics, ARO-MURI Quism program, Humboldt foundation. The work of I.M. was supported by the U.S. Department of Energy, Office of Science, Materials Sciences and Engineering Division. E.D. acknowledges hospitality of the Max Planck Institute for Quantum Optics, Ludwig Maximilian University, and Institute for Theoretical Studies  ETH, where part of this work has been completed.

%

\appendix

\section{Calculation of Noise}
\label{sec:noisecalc}

To the zeroth-order in the perturbative parameter $\Delta/\Delta_{0}$ ($\Delta$ is the gap between successive energy levels of the spin-pairs formed due to the neighbors of the spins in the pair forming the maximum gap $\Delta_0$), the noise contribution at every step of the RG is given by: 
\begin{align}
N(\omega) &= g^2 \sum_{m'} \abs{\matrixel{s,m}{\vec{A}}{s-1,m'}}^2 \delta(\omega + \omega_\downarrow) \nonumber \\
&+  g^2 \sum_{m'} \abs{\matrixel{s,m}{\vec{A}}{s+1,m'}}^2 \delta(\omega - \omega_\uparrow)
\end{align}

where $\vec{A} = \vec{S}_- - \frac{\vec{S}_-.\vec{S}_+}{\abs{\vec{S}_+}^2}\vec{S}_+$,  $g^2 = (g_A - g_B)^2/4$ is the geometrical factor proportional to the square of the difference of the coupling of the probe to the two spins $A$ and $B$ being combined at the given RG step, and $\omega_\uparrow$( $\omega_\downarrow$) is the energy difference between states of total angular momentum $s$ and $s+1$ ($s-1$) formed from the two spins. We will equivalently refer to these as the frequencies of the `up' and `down' transitions associated with a state with total momentum $s$. $m$ and $m'$ are the azimuthal quantum numbers associated with the states $s$ and $s+1$ (or $s-1$ in the second line). Note that there is no sum for $m$: all $-s < m < s$ contribute equivalently. Moreover, at this step of the RG, we only choose the magnitude $s$ of the effective spin, and not its completely quantum state.

The spectrum of two spins is readily given in terms of the coupling strength $J$ between the spins of size $s_A$ and $s_B$ being combined; the energy for such a spin pair combining into an effective spin of size $s$ is given by $E_s = \frac{J}{2} \left( s (s+1) - s_A (s_A + 1) - s_B (s_B + 1) \right)$. The energy difference between states $s$ and $s\pm 1 $ can be computed readily from this result. It is important to note that the sign of the Ferromagnetic and AntiFerromagnetic coupling implies a certain ordering of these levels, which changes the meaning of $\omega_\uparrow$ and $\omega_\downarrow$ accordingly. 

$\vec{S}_- = \vec{S}_A - \vec{S}_B $ is a rank $1$ spherical tensor operator (as one can ascertain from its commutation relations with the total spin angular momentum operator $\vec{S}_+$). Therefore, it only leads to transitions between states $s$ and $s \pm 1$, besides also having a matrix element that leaves $s$ unchanged.  The projection of $\vec{S}_-$ on to $\vec{S}_+$ (which is subtracted from $\vec{S}_-$ to yield $\vec{A}$) precisely removes this matrix element corresponding to transitions from $s \rightarrow s$; this projection operator is a low-frequency component of the two spins that must not be eliminated at the given RG step. 

The remaining aim of this sub-section is the calculation of the matrix element $\sum_{m'} \matrixel{s,m}{\vec{A}}{s \pm 1,m'} \matrixel{s \pm 1,m'}{\vec{A}}{s,m}$. 
By Wigner-Eckart theorem, the matrix elements are related to the Clebsch-Gordan coefficients as
$$\matrixel{s,m}{{A_q}}{s',m'} = c_{s, s'}<s,m||s',m'; 1,q>.$$
Thus, the up-transition rate is given by 
$$M(s,\uparrow)  = c_{s,s+1}^2\sum_{m',q} \abs{\matrixel{s,m}{}{s +1,m'; 1,q}}^2 = c_{s,s+1}^2,$$
and similar expression for $M(s,\downarrow) $.
Wigner-Eckart does not provide the values of the coefficients $c_{s, s'}$; they can be established, however in the following way. Consider the matrix elements of $A_0$, $\matrixel{s,m}{{A_0}}{s',m'} = c_{s, s'}<s,m||s',m'; 1,0>$ and $\matrixel{s',m'}{{A_0}}{s,m} = c_{s', s}<s',m'||s,m; 1,0>$ . They are obviously complex conjugates of each other. Thus, 

$$c_{s, s'} = c_{s', s}^*\frac{<s',m'||s,m; 1,0>}{<s,m||s',m'; 1,0>}  = c_{s', s}^*\frac{2s'+1}{2s+1},$$
where the last equality follows from the properties of 3$j$ symbols. On the other hand, 
\begin{align}
M(s) = M(s,\uparrow) &+ M(s,\downarrow) = 2 s_1 (s_1 + 1) + 2 s_2 (s_2 + 1) \nonumber \\
&- s( s + 1) - \frac{\left( s_1 (s_1 + 1) - s_2 (s_2 + 1) \right)^2}{s (s+1)}.
\end{align}
These two relationships are sufficient to calculate all transition rates separately. For the purposes of Sec.~\ref{sec:scalc}, it suffices to note that, generically, the sum $M(s)$ scales as $\sim s^2$, however, for $s = s_1 + s_2$ and $s = s_1 - s_2$, i.e., extremal states, as is the case in the $T = 0$ RG flow, $M(s = s_1 + s_2) = 4 s_1 s_2 / (s_1 + s_2 ) $ and $M(s = s_1 - s_2 ) = 4(s_1 + 1) s_2 / ( s_1 - s_2 + 1)$, implying, at $T = 0$, $M(s) \sim s$. 

\section{Analytical argument for scaling of the form factor at $T = \infty$}
\label{sec:noisefactinf}

In the main text, we argued that the probe form factor $F(q n_0 )\sim 1/qn_0$ for $q n_0 \gg 1$ in the infinite temperature case. Below we provide a more detailed justification of this result. 

Let us first recall that the probe form factor $F(q n_0)$ describes the scaling form of the squared difference of couplings $g_A$ and $g_B$, that is, $(g_A - g_B)^2/4$, when these individual couplings correspond to clusters whose size is $\sim n_0$ (note that the probe couplings contain the $q$-dependence). In this appendix, we show that this factor is related to the variance in the distribution of probe couplings which scales as $1/qn_0$ for $q n_0 \gg 1$.  

The general result for the effective coupling $g_+$ of the probe to a cluster formed by merging two clusters $A$ and $B$ of spins $\vec{S}_A$ and $\vec{S}_B$, coupled to the probe by strengths $g_A$ and $g_B$, is given by 

\be
g_+ = \frac{g_A + g_B}{2} + \frac{g_A - g_B}{2} \frac{\abs{\vec{S}_A}^2 - \abs{\vec{S}_B}^2}{\abs{\vec{S}_1 + \vec{S}_2}^2} .
\label{eq:gplus}
\ee

This expression in Eq.~(\ref{eq:gplus}) is valid for quantum spins; however, at the advanced stages of RG at $T = \infty$ we can treat them as classical variables. Then by averaging uniformly over spin directions, we can evaluate the average  $\avg{g_+}_{S_A,S_B} = (g_A + g_B) / 2$ and and the variance of $g_+$,  $\sigma^2(g_+)_{S_A,S_B} = \avg{g^2_+}_{S_A,S_B} - \avg{g_+}_{S_A,S_B}^2 = ( g_A - g_B )^2 / 4$. Here $\avg{...}_{S_A,S_B}$ implies averaging over spin orientations of the 2 constituent spins $\vec{S}_A$, $\vec{S}_B$ of a cluster; note that $g_A$ and $g_B$ will themselves have a distribution of values owing to the fact that they are themselves composed of smaller spin clusters. In what follows, we will use the notation $\avg{...}$ (without the subscripts) to imply the averaging over all internal spins of a cluster. From the above, we see that the noise factor $F(q \ell)$ obtained when combining two clusters of size $\ell$ with probe couplings $g_A$ and $g_B$ (which depend on $q$) is given precisely by the variance $\sigma^2(g_+)$. Thus, in order to find the noise factor $F(q \ell)$, we need to determine the distribution of $g_+$. 

The expression for $\avg{g_+}_{S_A,S_B}$ implies that the probe coupling to each cluster is approximately the average of the probe couplings to individual spins. Note that spins which form singlet pairs do not couple to the probe, and are not counted in this sum. However, as argued before, singlets are entropically unfavorable at $T = \infty$ and so clusters are primarily arrays of contiguous spins. Consequently, the $| \avg{g_+} | \sim \int_0^{\ell} dx \; \cos ( q x ) / \ell$. In particular, for $q \gg 1/\ell$, $| \avg{g_+} | \approx 1/q\ell$. 

Now we estimate the width $\avg{\sigma(g_+)} = \avg{( g^2_A + g^2_B - 2 g_A g_B ) / 4}$. We will self-consistently show that the factor $\avg{g_A g_B}$ can be dropped from this sum if these clusters have size $\ell \gg 1/q$. To this end, we first note that the phase difference between $g_A$ and $g_B$ is of the order of $e^{i q \ell}$ which is effectively random for $q \ell \gg 1$. Thus, such clusters have uncorrelated probe couplings and $\avg{g_A g_B} \approx \avg{g_A}\avg{g_B} \sim 1/(ql)^2$; the last relation is determined from the result of $\avg{g_+}$. Assuming that this term makes an unimportant contribution (in the large $\ell$ limit) to the variance, we find $\avg{\sigma(g_+)} = (\avg{g^2_A} + \avg{g^2_B} ) / 4$. If we merge clusters with similar lengths $\ell$ and similar values of  $\avg{g^2_A} \approx \avg{g^2_B} $, the resulting average value $\avg{\sigma^2( g_+)}$ goes down by a factor of 2; consequently, $\avg{\sigma^2( g_+)} \sim 1/\ell$ for $\ell \gg 1/q$. In particular, since the width of distribution begins to grow only for $\ell \gg 1/q$, $\sigma^2( g_+) \sim 1/q \ell$. This scaling self-consistently confirms our assumption to neglect the cross term $\avg{g_A}\avg{g_B} \sim 1/(ql)^2$ and completes the argument. 

Therefore, the probe form factor $F(q n_0)$, which is the scaling form of the factor $(g_A - g_B)^2/4$ when the size of clusters $A$ and $B$ is $\sim n_0$, scales as $\sim 1/(q n_0)$ for $q n_0 \gg 1$, as advertised in the main text.

\begin{center}
\begin{figure}[h]
\includegraphics[width=3in]{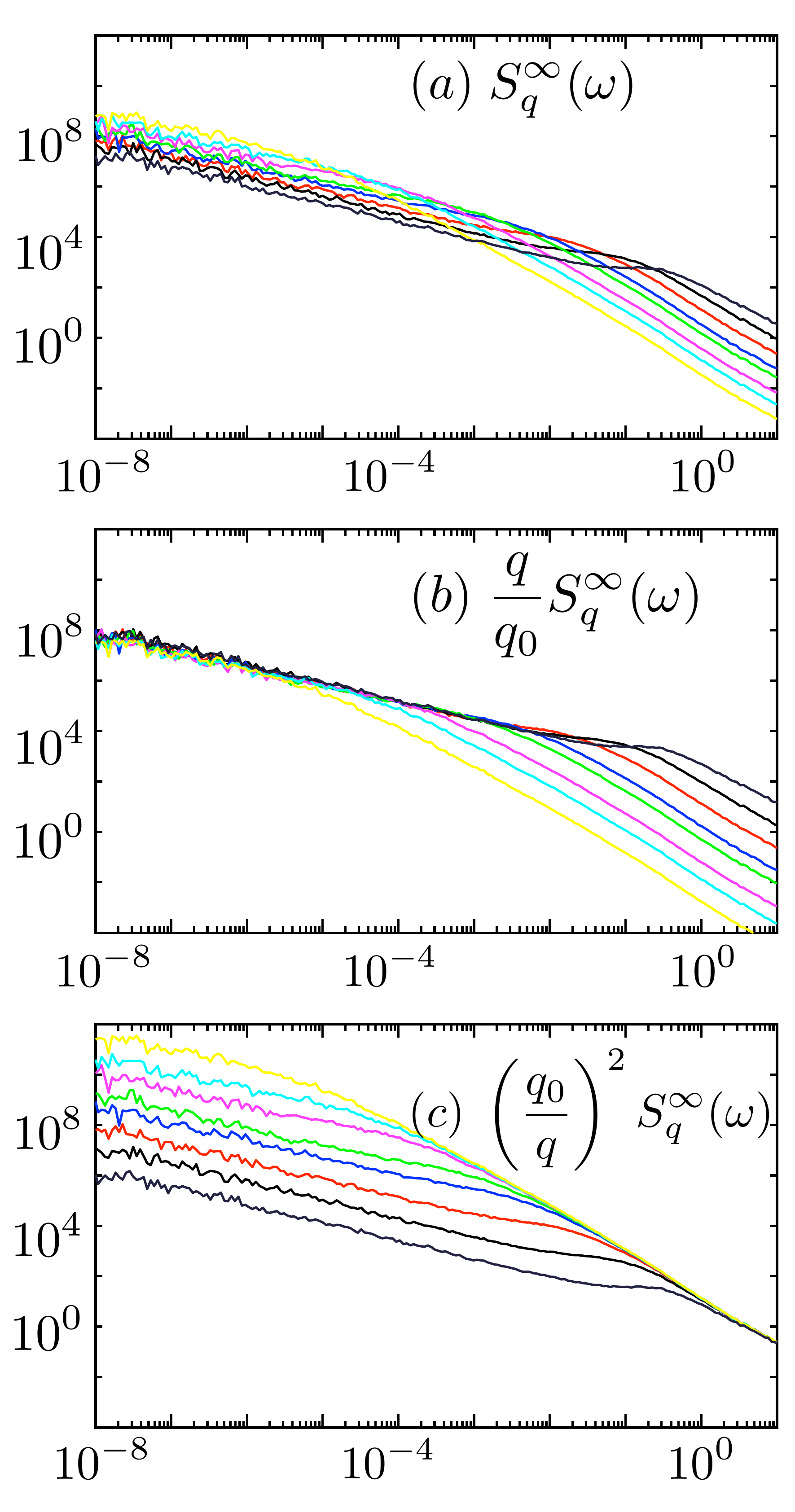}
\caption{(a) The $T= \infty$ dynamical structure factor $S^\infty_q(\omega)$ is plotted as a function of frequency for various $q = 2 \pi / l $ ($l \in [40, 800]$), system size $L = 15000$.  (b) $S^\infty_q (\omega) \sim 1/q$ for $\omega \ll q^{1/\beta}$ is evident from the scaling collapse at low frequencies. (c)$ S^\infty_q(\omega) \sim q^2 $ for $\omega \gg q^{1/\beta}$ is evident from the scaling collapse at high frequencies.}
\label{fig:qscalinginfT}
\end{figure}
\end{center}

\begin{center}
\begin{figure}[h]
\includegraphics[width=3in]{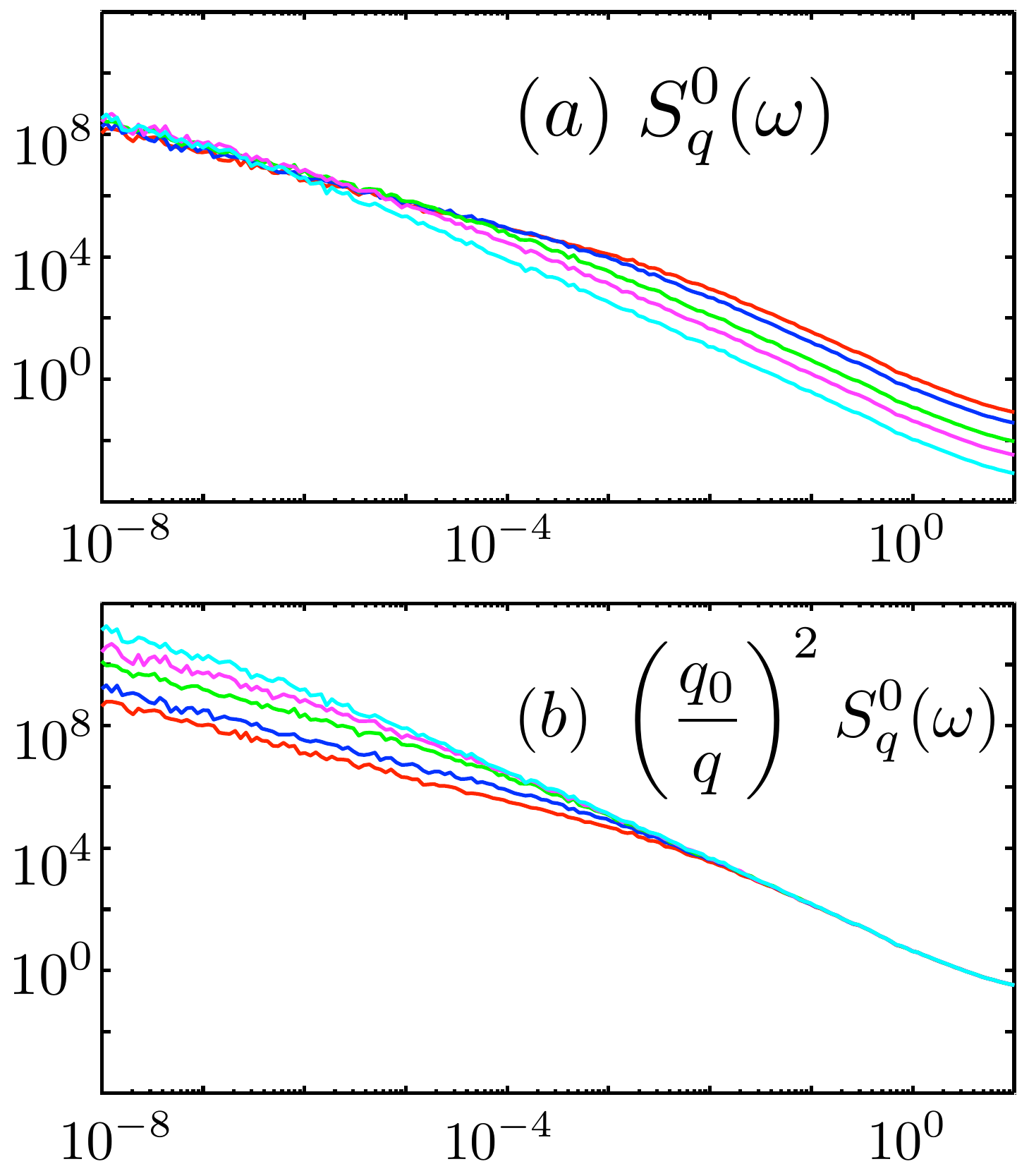}
\caption{(a) The $T= 0$ dynamical structure factor $S^0_q(\omega)$ is plotted as a function of frequency for various $q = 2 \pi / l $ ($l \in [40, 800]$), system size $L = 15000$. $S^0_q(\omega) \sim \text{const.}$ for $\omega \ll q^{1/\beta}$ is evident from the collapse at low frequencies (without any scaling). (b) $ S^0_q(\omega) \sim q^2 $ for $\omega \gg q^{1/\beta}$ is evident from the scaling collapse at high frequencies.}
\label{fig:qscalingzeroT}
\end{figure}
\end{center}

\section{$q$-dependent scaling of the Structure factor}
\label{sec:qscalingcheck}

Figs.~\ref{fig:qscalinginfT} and~\ref{fig:qscalingzeroT} detail the $q-$dependent behavior of the structure factor $S^\infty_q (\omega)$ at $T = \infty $ and $S^0_q (\omega)$ at $T = 0 $ respectively, at high ($\omega \gg q^{1/\beta}$) and low ($\omega \ll q^{1/\beta}$) frequencies. The $q-$dependence is in accordance with the results in the main text. 

\section{Comparison of $T = 0$ and $T = \infty$ noise:}
\label{sec:tzeroinf}

\begin{center}
\begin{figure}[h]
\includegraphics[width=3in]{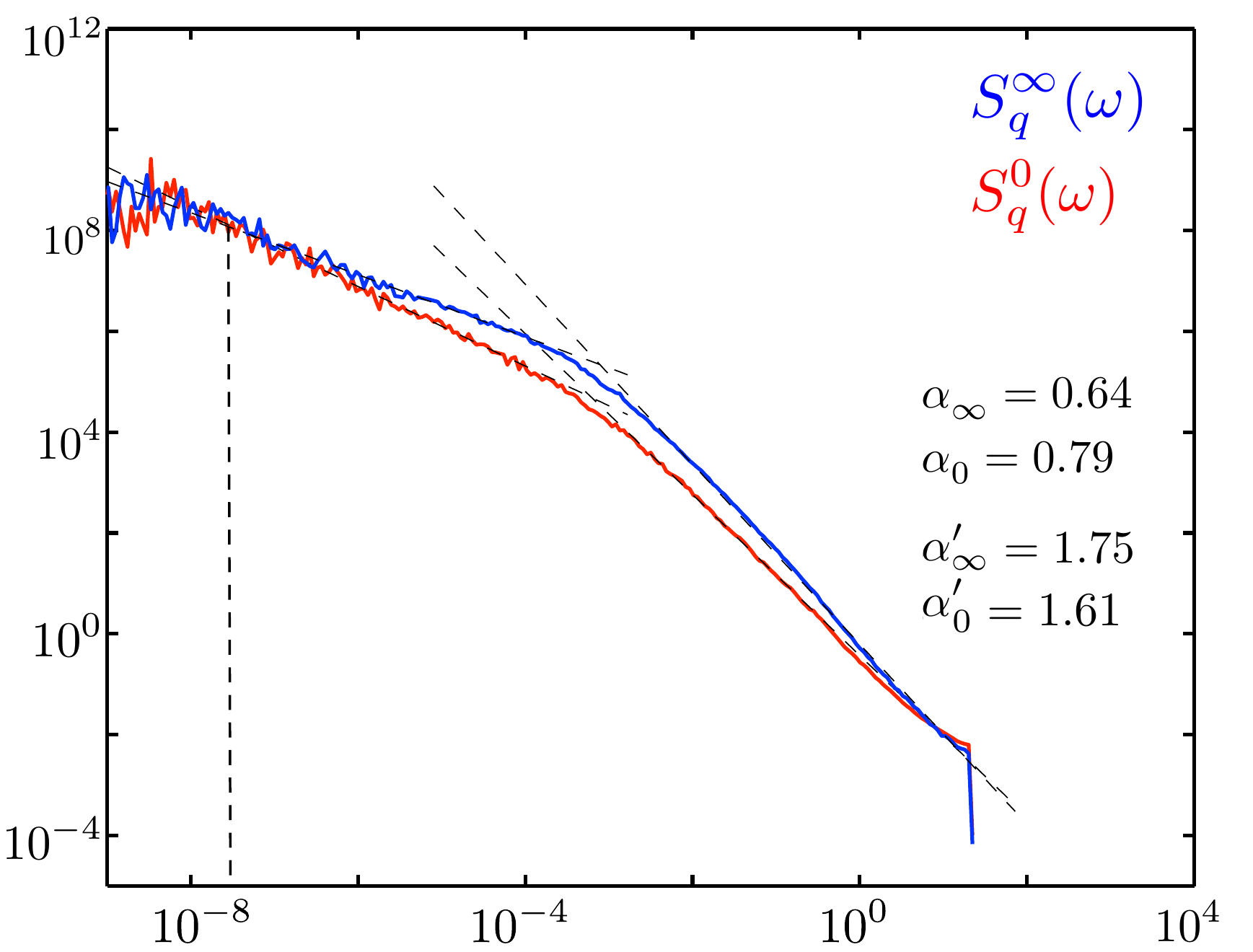}
\caption{The structure factors $S^\infty_q (\omega)$ ($T = \infty$, blue) and $S^0_q (\omega)$ ($T = 0$, red) are plotted simultaneously, for initial bias $\eta_i = 0$, disorder strength $D = 3$, system size $L = 15000$, and $q = 2 \pi / 160$. The powers $\alpha'$ and $\alpha$ are also indicated for the two cases with the corresponding subscripts. They approximately satisfy the respective scaling relations for zero and infinite temperature cases.}
\label{fig:zeroinfTcompare}
\end{figure}
\end{center}

Fig.~\ref{fig:zeroinfTcompare} shows a direct comparison of zero and infinite temperature noise for single disorder realization. There are three aspects of this plot that are of significance: i) the magnitude of the noise does not vary much with temperature, while the power laws change slightly; ii) (for the case of a 50-50 mix of F/AF bonds) the low frequency power law seems to increase with decreasing temperature, and; iii) the noise magnitude seems to be fairly constant at approximately $\omega = 10^{-8}$. A curious observation in nearly all experiments is that the noise magnitude does not change at approximately $1 Hz$, which is about a factor of $10^{-9}$ smaller than the typical microscopic interaction strength amongst the spins. In contrast, in our simulations, the microscopic scale is $e^D \sim 10$, $D = 3$, and noise magnitude is observed to be roughly constant at about $\omega \approx 10^{-9} e^D$. 

We note that the above comparison is based on the assumption that the two-dimensional network of spins producing flux noise in SQUIDs is likely more tightly confined in one direction (along the width of the conducting strip) than the other (along the length of the conducting strip), and that this implies that the behavior of the spin-network mirrors that of one-dimensional spins below a certain frequency scale. 

\section{Numerical checks / convergence of RG}
\label{sec:numchecks}

\begin{center}
\begin{figure}[h]
\includegraphics[width=3in]{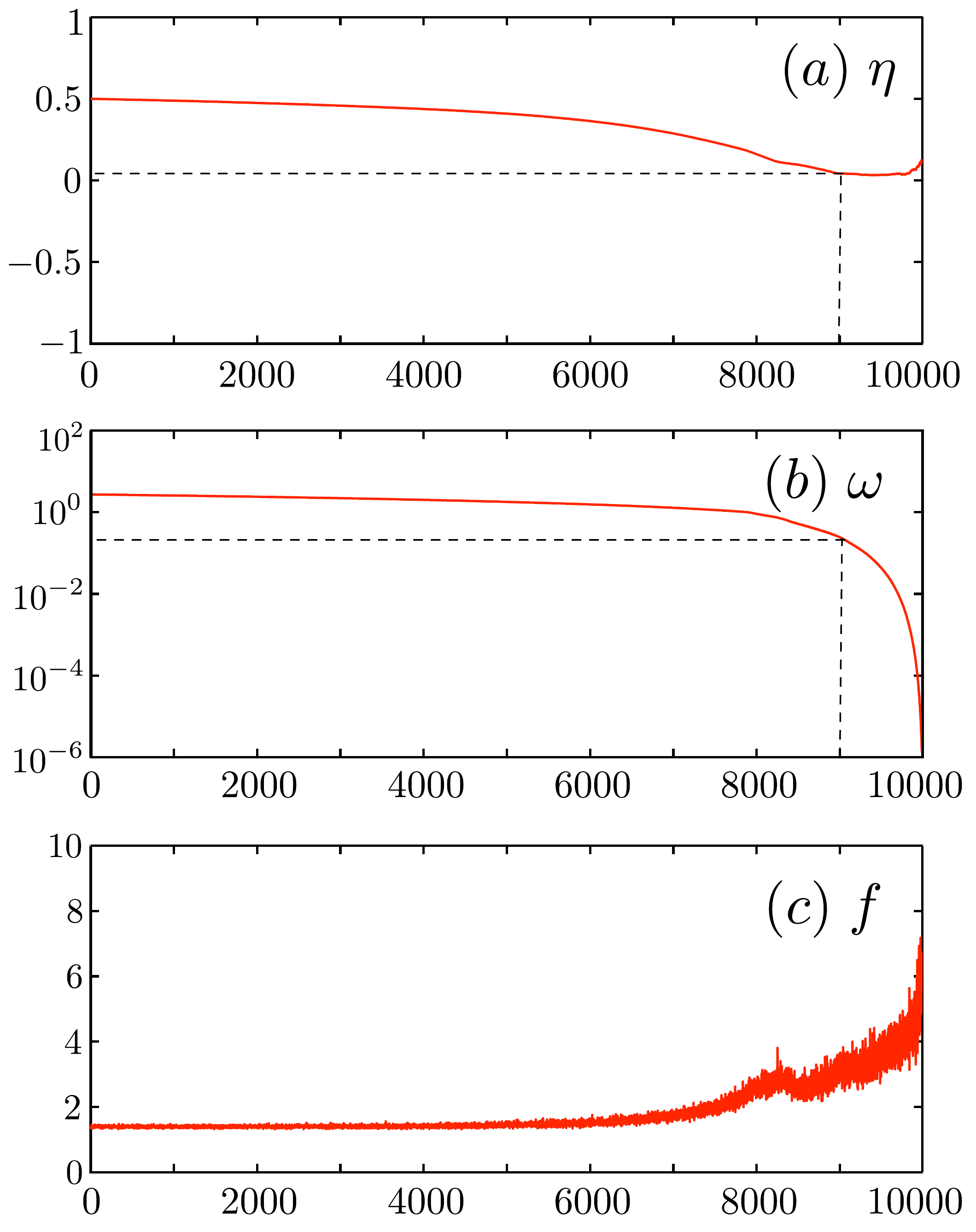}
\caption{The average (a) bias, (b) noise frequency and (c) `fidelity' is plotted against the RG step for a system of size $L = 10000$. Initial bias $\eta_i = 0.5$, temperature $T  = 0$,  disorder strength $D = 1$.}
\label{fig:figRGconvergence}
\end{figure}
\end{center}

\begin{center}
\begin{figure}[t]
\includegraphics[width = 2in]{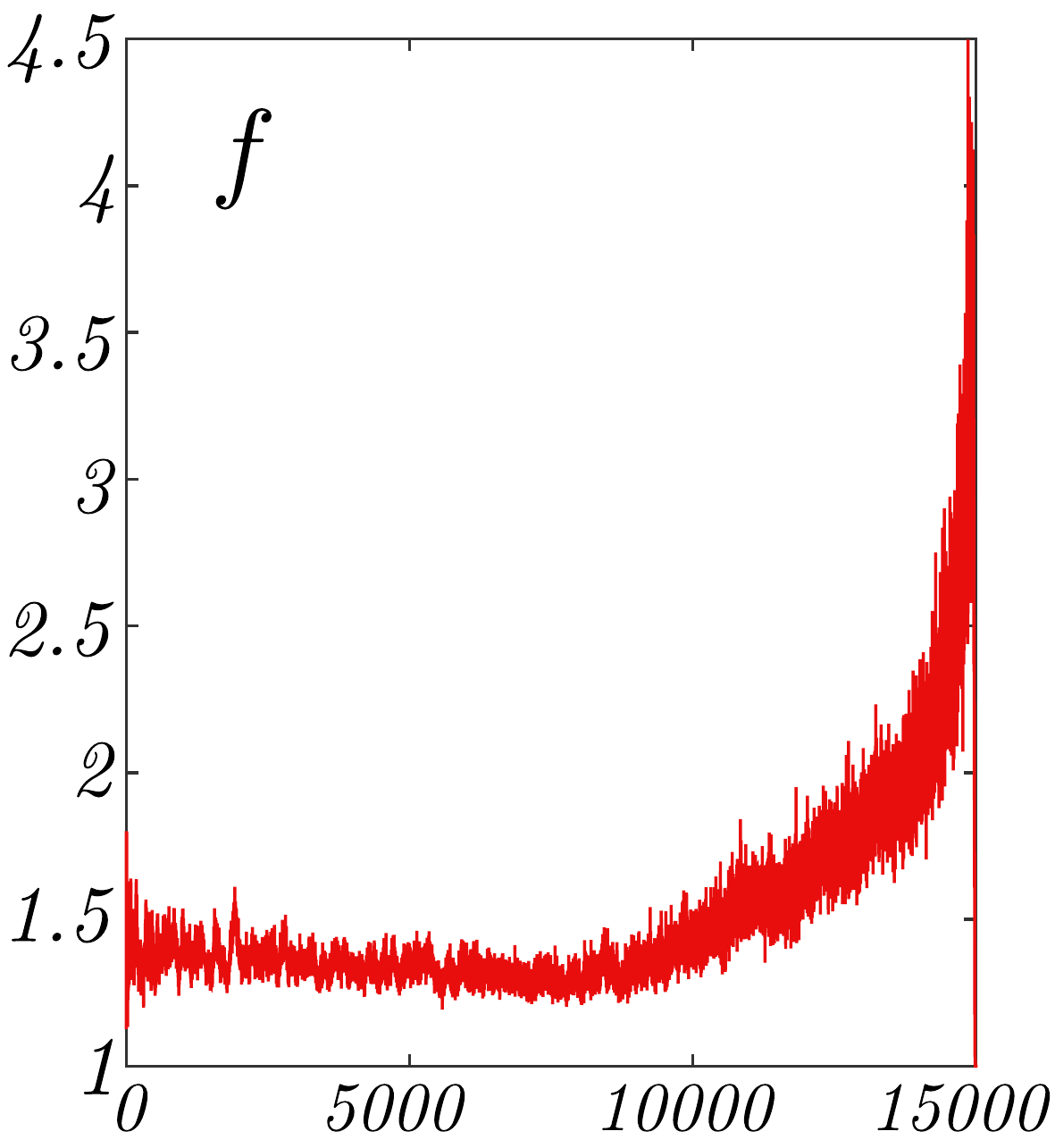}
\caption{The fidelity is plotted as a function of RG steps at $T= \infty$ for a system of size $L = 15000$ and initial disorder strength $D = 1$. }
\label{fig:fidelityinfT}
\end{figure}
\end{center}

Fig.~\ref{fig:figRGconvergence} plots (a) bias, (b) noise frequency and (c) `fidelity' is plotted against the RG step for a system with a given set of initial conditions (bias, disorder strength, system size, temperature). Although plotted for a specific system, the following observations are fairly independent of the various system parameters, including temperature. The bias is seen to flow to a a stable value after about 90\% of the initial spins have been eliminated. This value is close to the 50\% - 65\% ($\eta = 0$ - $0.3$) mark although it has small variations which depend on the initial bias of the system at $T = 0$, but not at $T = \infty$. The noise frequency (of eliminated spin-pair) dips below microscopic values only after these initial eliminations. These observations suggest that the RG converges and the (high- and low-) frequency dependent noise observed at both ends of the crossover ($\omega \sim q^{1/\beta}$) is from the limit where the RG has converged. The fidelity plotted is the log-averaged (typical) estimate of the inverse of the perturbation parameter, that is, $f = \Delta_0/\Delta$, where $\Delta_0$ is the gap due to the spin-pair being eliminated at the given RG step, and $\Delta$ is the maximum gap formed by the coupling of one of these spins (in the spin-pair) to its neighbors.  It is seen to rise dramatically as the RG reaches convergence indicating that the RG is fairly successful; since the fidelity is large but finite we conclude that the RG flows to a strong-randomness fixed point, and not an infinite-randomness point. In Fig.~\ref{fig:fidelityinfT}, the fidelity is plotted for the infinite temperature RG simulation and is seen to also perform quite well. Note that, at zero temperature, the fidelity is best for the initially completely AF system, and becomes progressively worse with the addition of more F bonds. 

\begin{center}
\begin{figure}[h]
\includegraphics[width=3in]{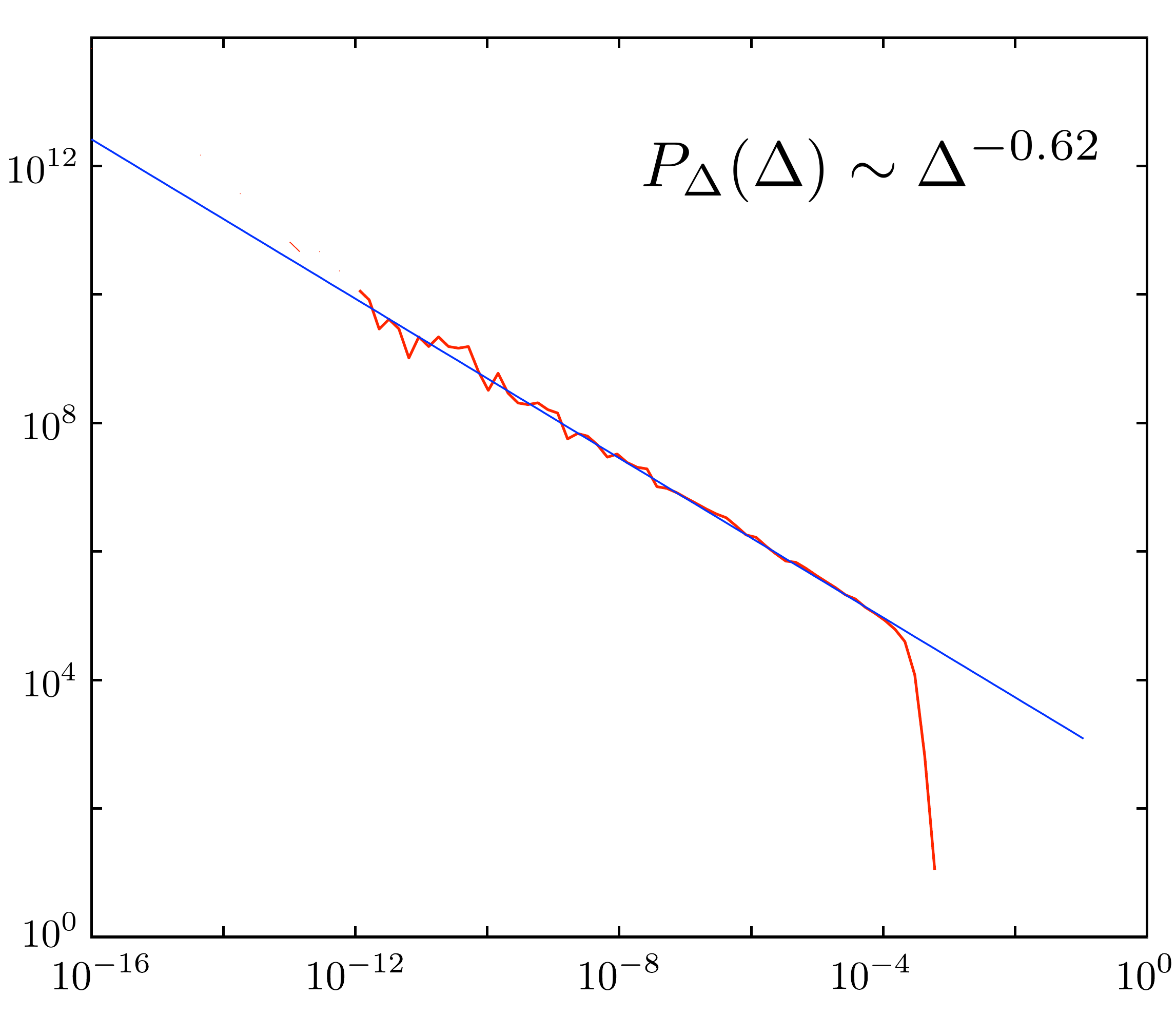}
\caption{Distribution of the gap $\Delta$ after 98\% of the spins were eliminated. The system has Initial bias $\eta_i = 0.5$, temperature $T  = 0$,  disorder strength $D = 1$. The noise power law $\alpha$ and the power law of the distribution of the gap are not the same.}
\label{fig:powerlaw}
\end{figure}
\end{center}

Fig.~\ref{fig:powerlaw} plots the distribution of the gap $\Delta$ after 98\% of the spins were eliminated for the same system as above. The power law form of the distribution guarantees that the typical gaps are much smaller than the largest gap in the system, which ensures that the RG works (and increases the fidelity). The noise power law $\alpha$ and the power law of the distribution of the gap is not the same generally. The the purely AF case, the distribution is $D_\Delta(\Delta) \sim 1/\Delta$, as is known from the analysis of the IRFP in the AF system at $T = 0$. As the initial distribution becomes more ferromagnetic, the exponent of the power law approaches $0$. The smaller exponent also marks a reduction in the fidelity and the efficacy of the RG. 

\begin{center}
\begin{figure}[h]
\includegraphics[width=3in]{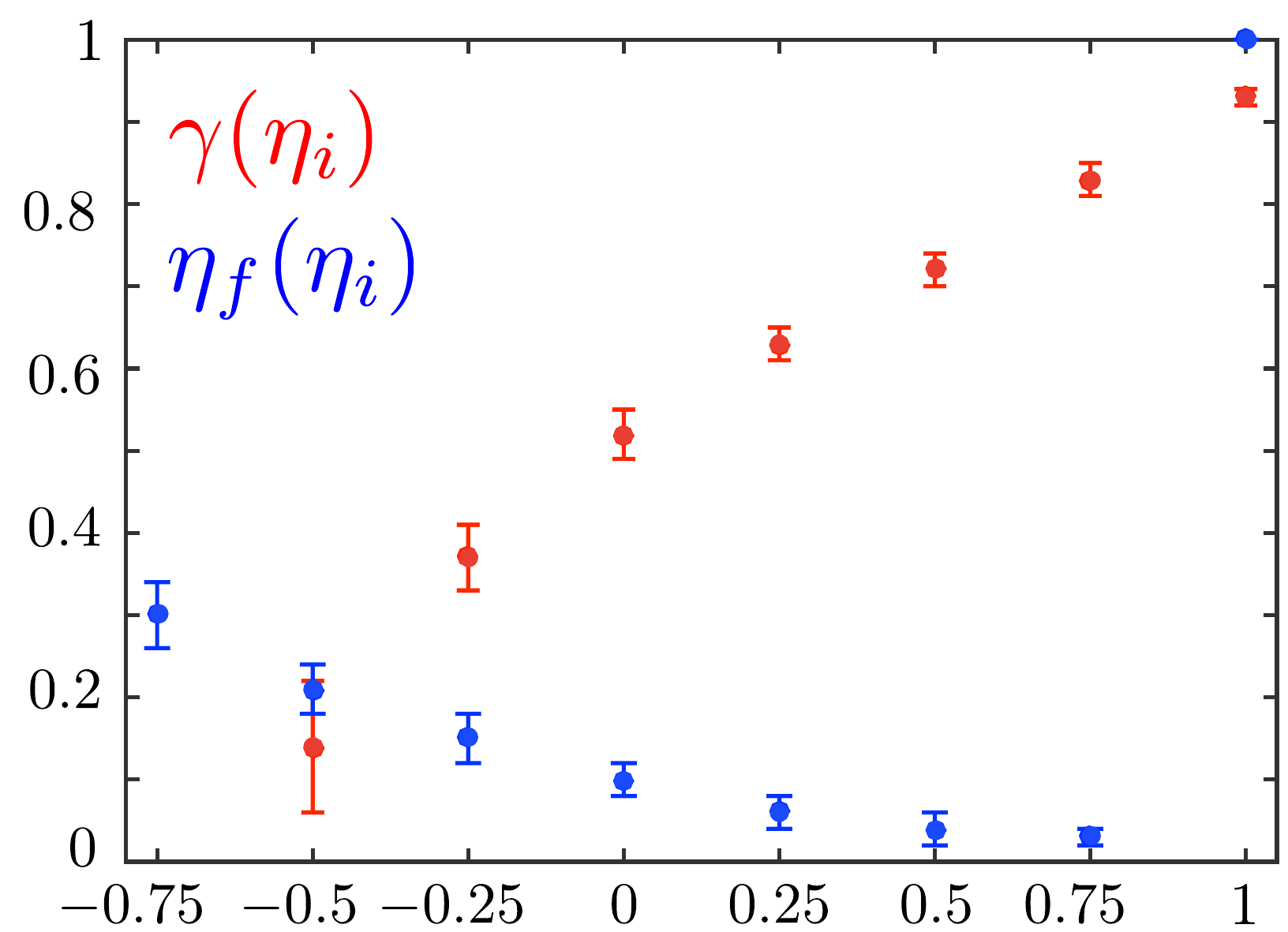}
\caption{The power $\gamma$ (red) of the gap distribution $P_\Delta (\Delta) \sim 1/\Delta^\gamma$ and the final bias $\eta_f$ (blue) are plotted as a function of the initial bias $\eta_i$ of a $1/|J|$ distribution with disorder strength $D = 3$.}
\label{fig:rgres}
\end{figure}
\end{center}

The power $\gamma$ of the gap distribution $P_\Delta (\Delta) \sim 1/\Delta^\gamma$ and the final bias $\eta_f$ are plotted as a function of the initial bias $\eta_i$ of a $1/|J|$ distribution with disorder strength $D = 3$ in Fig.~\ref{fig:rgres}.

\section{Relation between power laws of the gap distribution and the noise exponent}
\label{sec:exprelate}

The power laws of the gap distribution and the noise exponent are related. This can be seen as follows. The number of bonds $N_t$ at a given maximum gap value $\Delta_0$ is given via the dynamical exponent $z$ as $N_t (\Delta_0) \sim \Delta_0^{1/z}$. Some of these bonds are F and some AF; we assume that the proportion F/AF bonds at the maximum gap $\Delta_0$ is $p_{F/AF} (\Delta_0)$. We assume a power-law form of the distribution of the F/AF bonds associated with gaps $\Delta$ given by $Q_{F/AF} (\Delta, \Delta_0) = \frac{(1- \gamma_{F/AF})}{\Delta_0} \left( \frac{\Delta_0}{\Delta} \right)^{\gamma_{F/AF}}$. In the regime that the RG has converged, the exponents $\gamma_{F/AF}$ and proportions $p_{F/AF}$ become independent of the maximum gap $\Delta_0$. 

We now prove the relation $1/z = p_F (1 - \gamma_F) + p_{AF} (1 - \gamma_{AF})$ as follows. The number of F/AF bonds eliminated at every step of the RG is given by $ d N_{F/AF} (\Delta_0) = N_t (\Delta_0) Q_{F/AF} (\Delta_0, \Delta_0) d \Delta_0 $. The sum of the eliminated F/AF bonds gives the number of total eliminated bonds, which can alternatively be determined through the dynamical exponent as $ d N_t (\Delta_0) = \frac{1}{z} \frac{d \Delta_0}{\Delta_0}$. A comparison of the results obtained yields the aforementioned relation. Note also that the fact that the proportion of F/AF bonds becomes constant implies that $d N_F (\Delta_0) / d N_{AF} (\Delta_0) = p_F / p_{AF}$. This further implies that $\gamma_F = \gamma_{AF}$. 

The above results can also be used to determined the noise exponent via the usual relations $\alpha = 1 - 1/2z$ (at $T = 0$) and $\alpha = 1 - 1/z$ (at $ T = \infty$). These relations are discussed in Sec.~\ref{sec:scalingapproach} In particular, for the case of the disordered Heisenberg AF chain, we know that $p_{F} = 0$, $p_{AF} = 1$, $\gamma_{AF} = 1$, and consequently, $\alpha = 1$.

\section{Distribution $D(n)$ of cluster size at $T  =0$}
\label{sec:dnform}

\begin{center}
\begin{figure}
\includegraphics[width=3in]{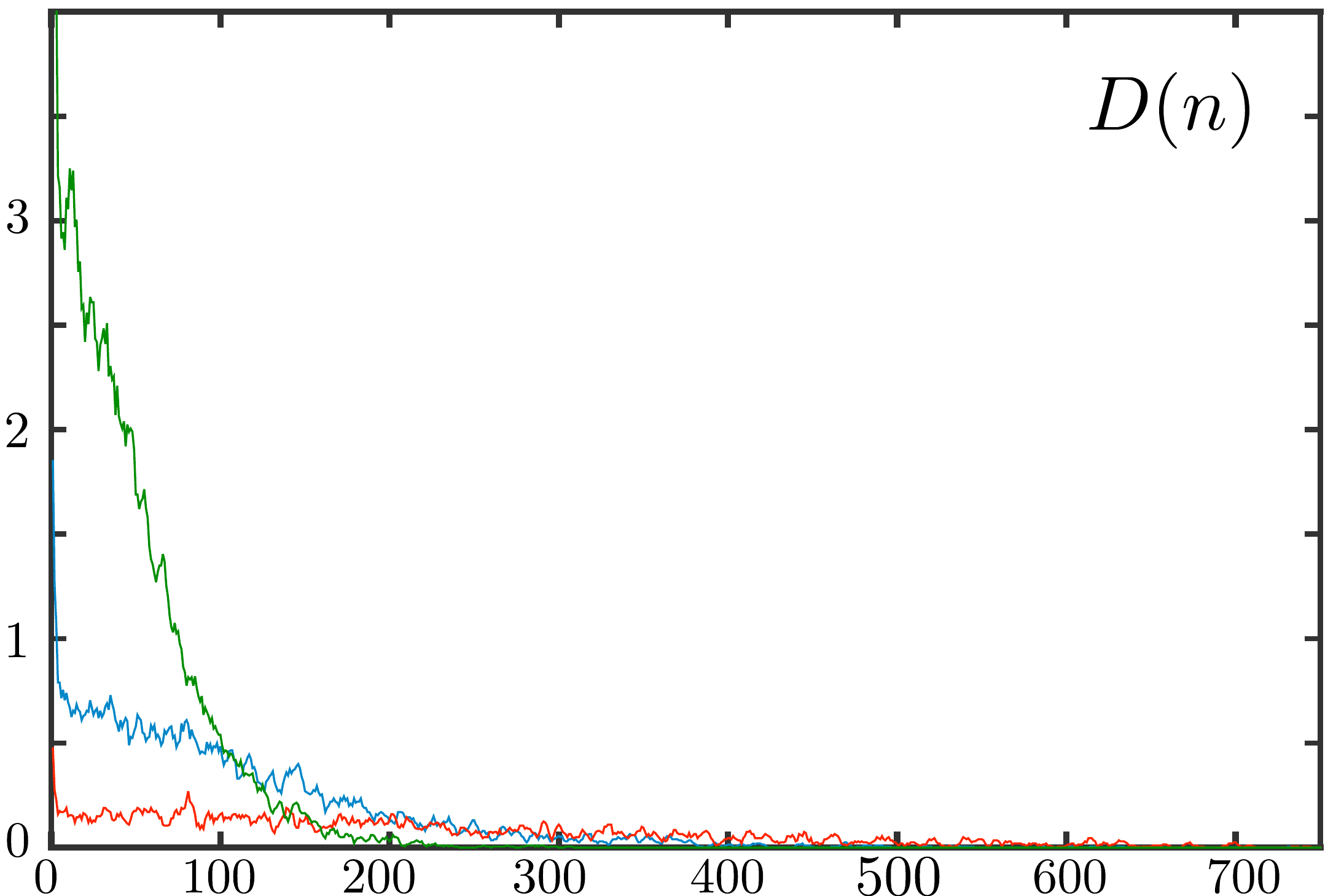}
\caption{The distribution $D(n)$ of the cluster sizes $n$ at various (different colors) stages of the RG for $L = 15000$ spin chain at $T = 0$, and initial bias $\eta_i  = 0.5$. The features of the result are not dependent on the specifics of the system parameters. The important facet here is that $D(n)$ has appreciable weight at all stages of the RG in the $n \rightarrow 0$, while it decays beyond a certain scale $n_0$ which depends on the energy scale at which the RG was terminated. This justifies the form of the distribution we use in the main text to calculate the $T =0 $ finite-$q$ susceptibility of the spin chain.  }
\label{fig:dnfig}
\end{figure}
\end{center}

It was mentioned in the main text that the distribution of clusters at $T = 0$ follows a form $D(n) \sim 1/n_0$ for $n \lesssim n_0$ and small for $n \gtrsim n_0$, where $n_0$ is the typical cluster size at any RG scale. Such a distribution was shown to lead to the failure of linear response theory at $T  = 0$, primarily because it guarantees the presence of clusters of small sizes (and especially, of the probe wavelength $2\pi/q$) at all energy scales of the RG; these small clusters were found to give a divergent contribution to the susceptibility of the system even at finite-$q$. In Fig.~\ref{fig:dnfig} we justify this form of the distribution function numerically.

\section{Correlations between the coupling and the spin size}
\label{sec:jscorr}

\begin{center}
\begin{figure}
\includegraphics[width=3in]{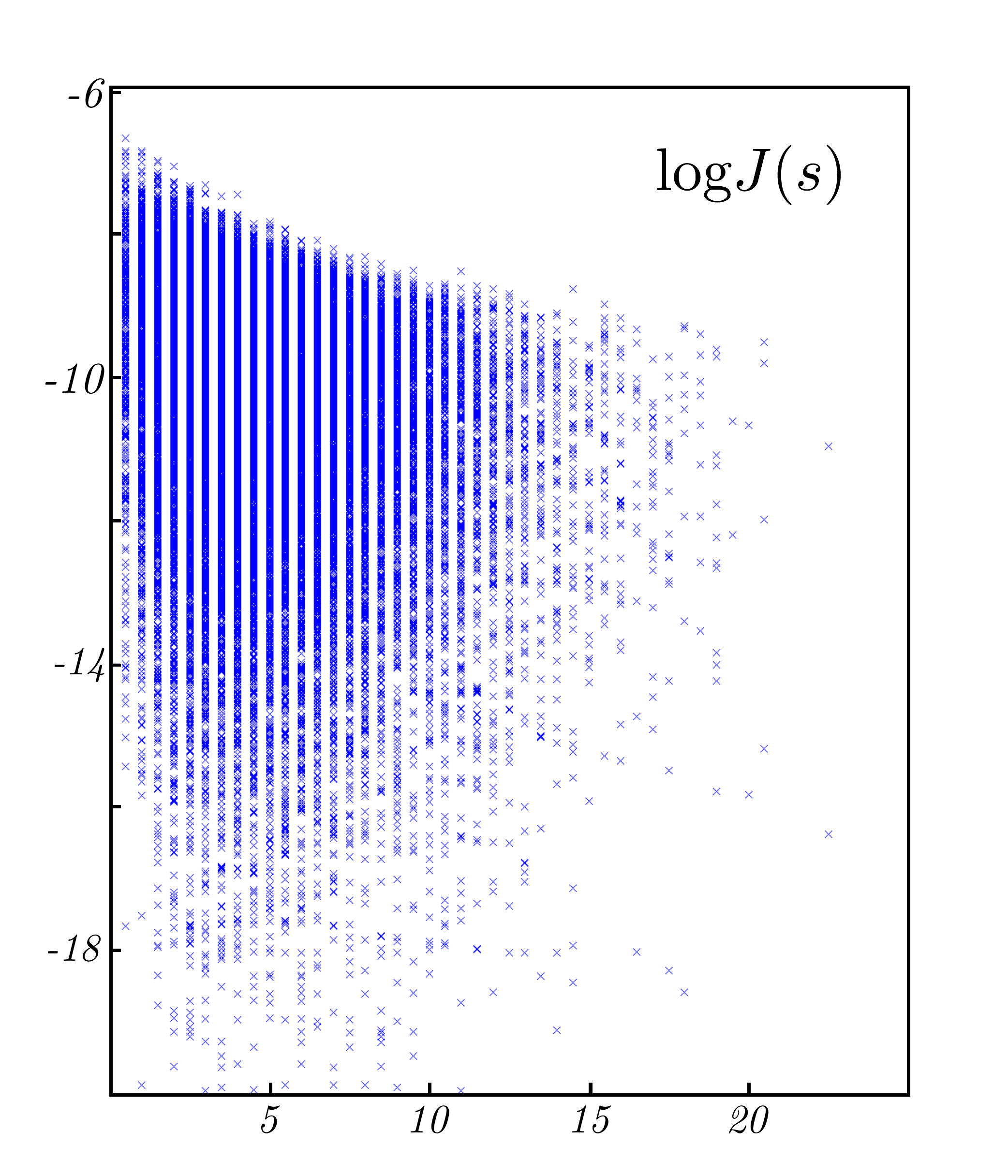}
\caption{Scatter-plot of the logarithm of the couplings $log(J)$ of effective spins of size $s$ remaining after the RG procedure has reached convergence operating on a $L = 5000$ spin chain at $T = \infty$. The maximum coupling is seen to decrease with the spin size $s$}
\label{fig:jscorr}
\end{figure}
\end{center}

It was argued in Ref.~\cite{vasseur2014dynamics} that the RG procedure fails because it generates exceedingly large spins. This conclusion is incorrect; it is important to note that these larger spins also couple much more weakly to their neighbors, as seen in Fig.~\ref{fig:jscorr}, and that this balance precisely ensures that the RG can flow to a strong-randomess fixed point with tractable scaling properties. The authors of Ref.~\cite{vasseur2014dynamics} perform RSRG by keeping the length of the chain fixed and attaching new spins with arbitrary sizes but also arbitrary coupling strengths which misses out on these correlations. Moreover, they perform their RG by eliminating the largest bond value $J$ rather than the larger gap $\Delta$.

\section{Simulations in two dimensions}
\label{sec:2dsimul}

\begin{center}
\begin{figure}
\includegraphics[width=3in]{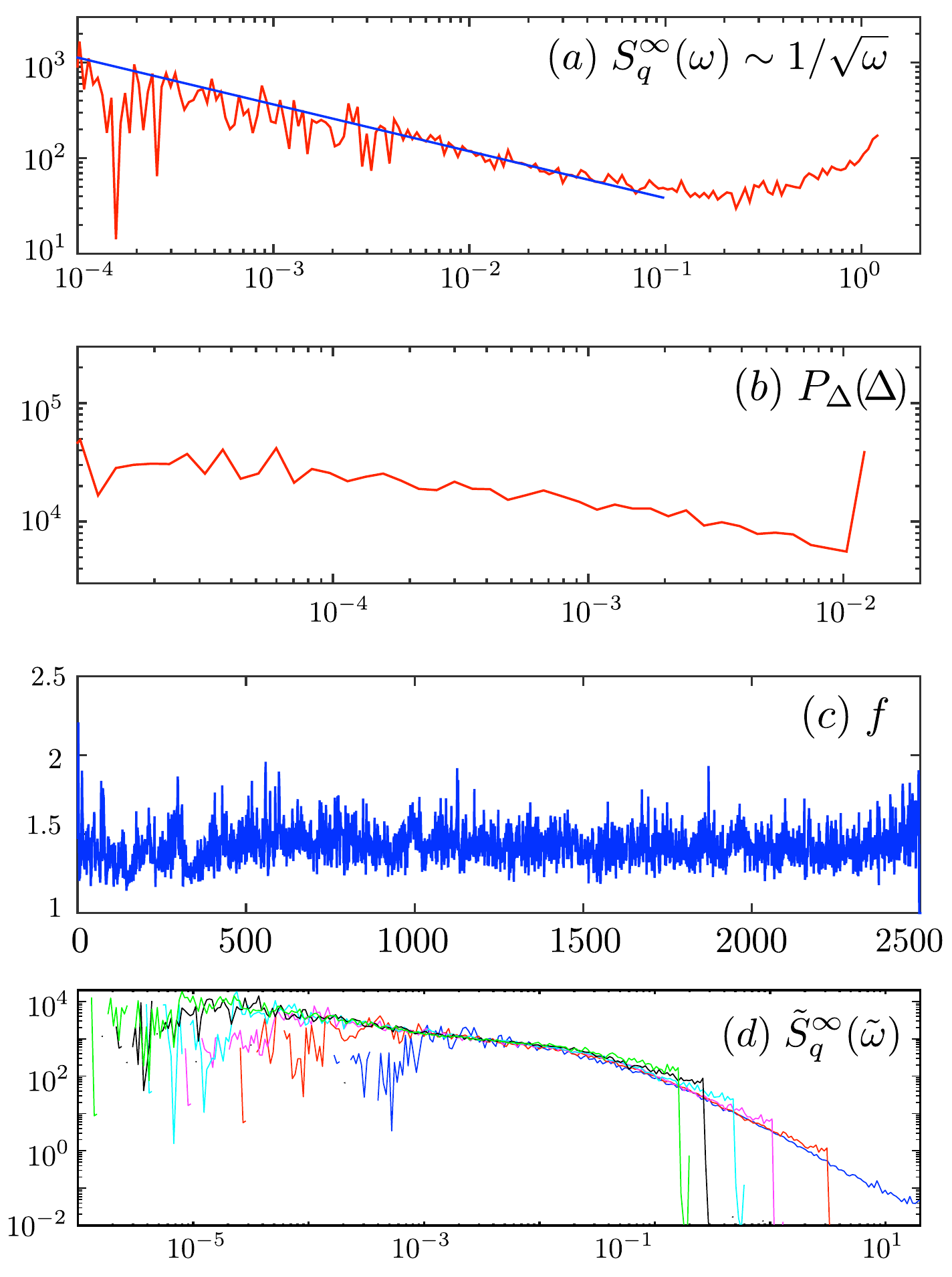}
\caption{Various features of the simulations on two-dimensional system of size $50 \; \times \; 50$ are shown. (a) The noise spectrum for low frequencies has a power law tail $S_q (\omega) \sim 1/\sqrt{\omega}$ ($q = (\pi, \pi)$ here); (b) The distribution of gaps is fairly broad; (c) The fidelity $f$ does not rise over the course of RG; (d) Approximate scaling collapse using the Generalized-diffusion ansatz using exponent $\beta = 0.55 \pm 0.05$ for wave-vectors $q = (2 \pi n / L, 2 \pi n / L)$ with $n \in [2 (\text{blue}),4 (\text{red}), 6(\text{magenta}), 8(\text{cyan}), 10(\text{black}), 12(\text{green})]$. and $L = 50$. The lower (upper) frequency power laws are found to be: $\alpha \approx 0.5 \pm 0.05$, $\alpha' \approx 1.75 \pm 0.05$; it is unclear whether there is data for a sufficiently large range of frequencies to extract the high-frequency power law reliably, especially since it exhibits a drift to higher values as we go to smaller wave-vectors.}
\label{fig:2dfig}
\end{figure}
\end{center}

\begin{center}
\begin{figure}
\includegraphics[width=3in]{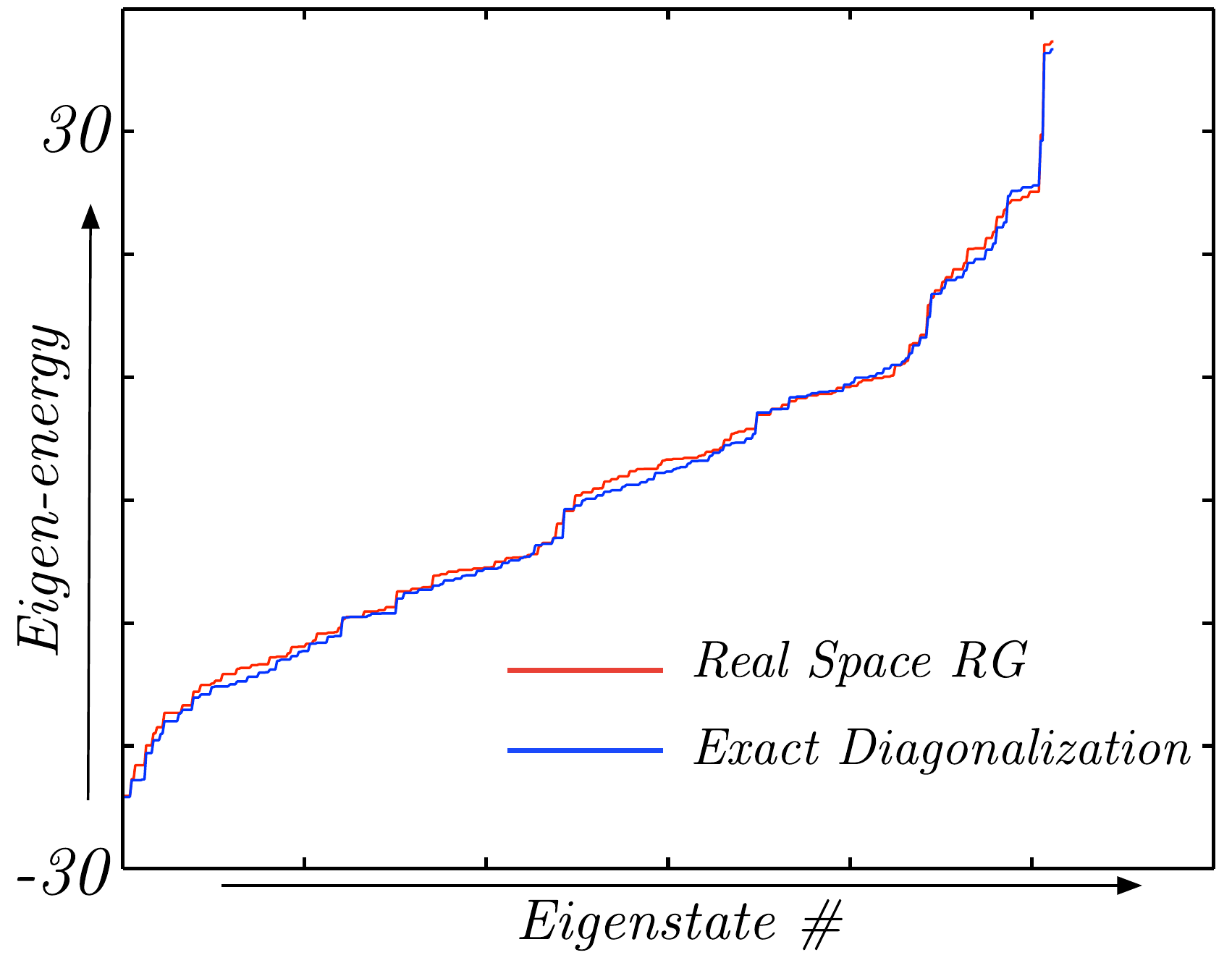}
\caption{The energy spectrum of a small $3 \; \times \; 3$ lattice spin system (disorder strength $D = 3$) is calculated using exact diagonalization (blue) and compared with the RG generated spectrum (red).}
\label{fig:2drged}
\end{figure}
\end{center}

\begin{center}
\begin{figure}
\includegraphics[width=3in]{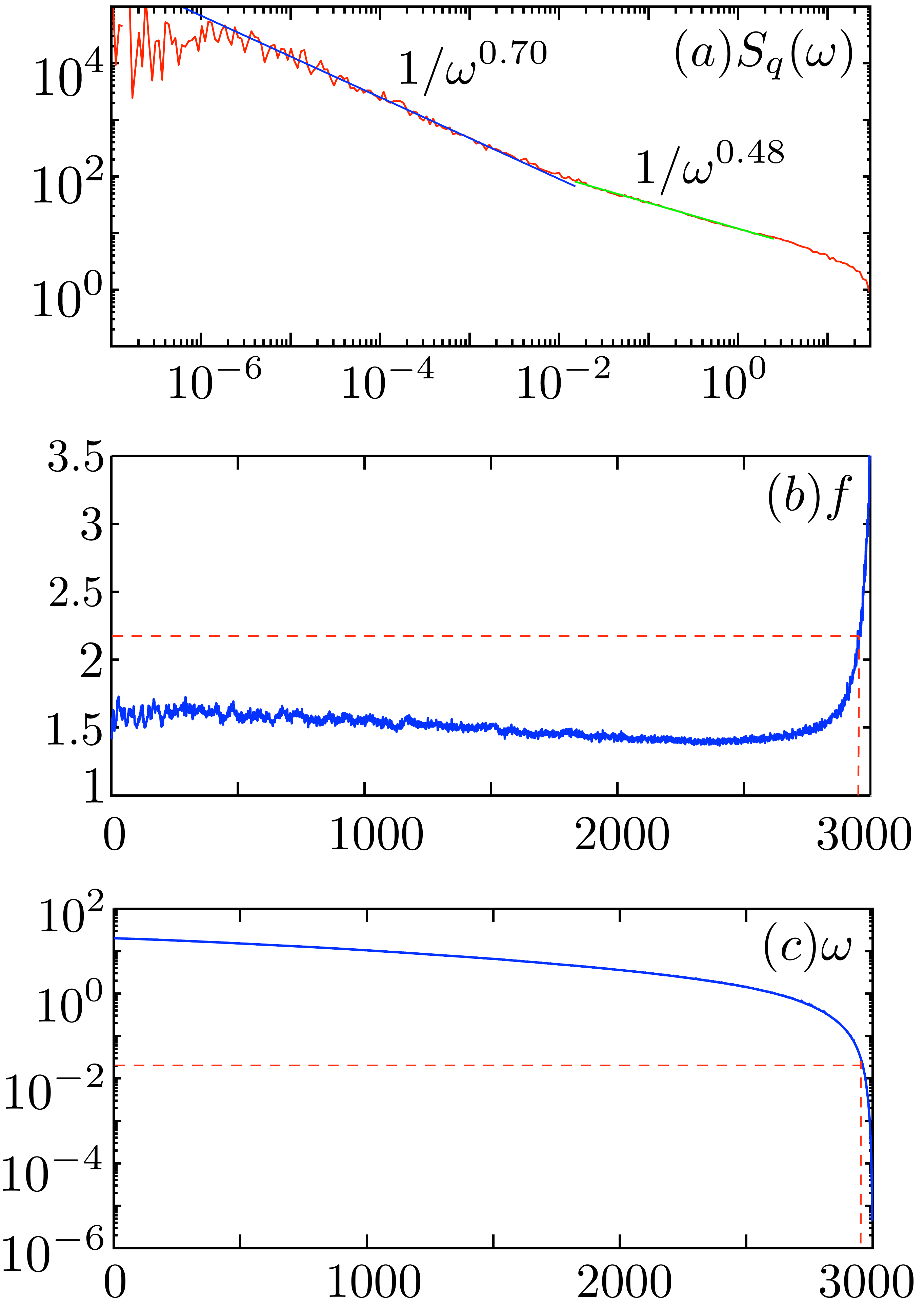}
\caption{(a) The finite-$q$ structure factor $S_q (\omega)$ [$q = (2 \pi / 25,\pi$)] as a function of frequency; (b) the fidelity $f$ and (c) the noise frequency are plotted as a function of the RG steps for a two-dimensional strip of spins of lattice size $500 \; \times \; 6$, with $D = 3$ at $T = \infty$. There are $\sim 50$ clusters remaining when these are of the size of the width of the spin network; these remaining clusters behave as a one-dimensional network of effective spins and produce the anomalous $1/\omega^\alpha$ noise. Here we see that about $50$ (red dashed lines) spins remain when the maximum frequency $\omega \sim 2 \times 10^{-2}$ and this approximately agrees with the frequency below which $1/\omega^\alpha$ noise behavior is observed. Above this frequency, a $\sim 1/\sqrt{\omega}$ tail is observed as in the case of the square network of spins.}
\label{fig:2dstrip}
\end{figure}
\end{center}

The basic features of the results in 2D (at $T = \infty$) are outlined in Fig.~\ref{fig:2dfig}. The simulations (square geometry) were carried out for a $50 \; \times \; 50$ lattice of spins. The fidelity does not improve markedly over the course of the RG. However, the computed structure factor shows approximate scaling collapse according to the Generalized-diffusion form of the structure factor with an inverse dynamical exponent $\beta \approx 0.55 \pm 0.05$.  

While the RG doesn't successfully converge for large systems with a square geometry, it still manages to recover the full energy spectrum for sufficiently disordered, small systems with reasonable success (see Fig.~\ref{fig:2drged}). To simulate a two-dimension strip of spins, simulations were carried out for a $500 \; \times \; 6$ lattice of spins. Below a certain frequency scale, clusters become of the size of the width of the network, and the $1/\omega^\alpha$ noise behavior of one-dimensional systems with an anomalous exponent $\alpha$ is recovered (see Fig.~\ref{fig:2dstrip}). For a disorder strength of $D = 3$, $\alpha \approx 0.7$ was found.

\section{Role of faraway resonances}
\label{sec:roleofresonances}
Here we present an argument (based on a generalization of the analysis in the supplementary of Ref.~\cite{vosk2013dynamicalrg}) that suggests that once the system develops an dynamical exponent $z = 1/\beta > 1$, faraway resonances can be neglected. On the basis of the numerical evidence, we assume a power-law distribution of gaps $D(\Delta) \sim \frac{1-\gamma}{\Delta_0} \left(\frac{\Delta_0}{\Delta} \right)^\gamma$. Let us examine the possibility that two faraway spin-pairs, with local gaps separated by $d \Delta$ become resonant with one-another. The density of spin-pairs separated by a local gap $d \Delta$ is $\rho = D(\Delta_0) d \Delta = (1- \gamma) (d \Delta / \Delta_0)$ and the typical distance between such approximately-resonant spin-pairs is $ l = 1/\rho$. We would like to compare $d \Delta_0$ to the effective coupling between these spin-pairs. Since there is a fixed dynamical exponent $1/\beta$ in our system, this effective interaction between the spin-pairs is given by $\Delta_{int} = l^{-1/\beta}$, where $l$ is again the typical distance found above. 

For faraway resonances to be unimportant, we require $d \Delta \gg \Delta_{int} (d \Delta) =$ const.$\; (d \Delta) ^{1/\beta}$. This condition is always satisfied if $\beta  < 1$, a condition which is fulfilled in all our simulations. Thus, we expect resonances to be unimportant, so long as the RG generates a power-law distribution of gaps, and the dynamical exponent $z = 1/\beta > 1$. 

\begin{center}
\begin{figure}[h]
\includegraphics[width=3in]{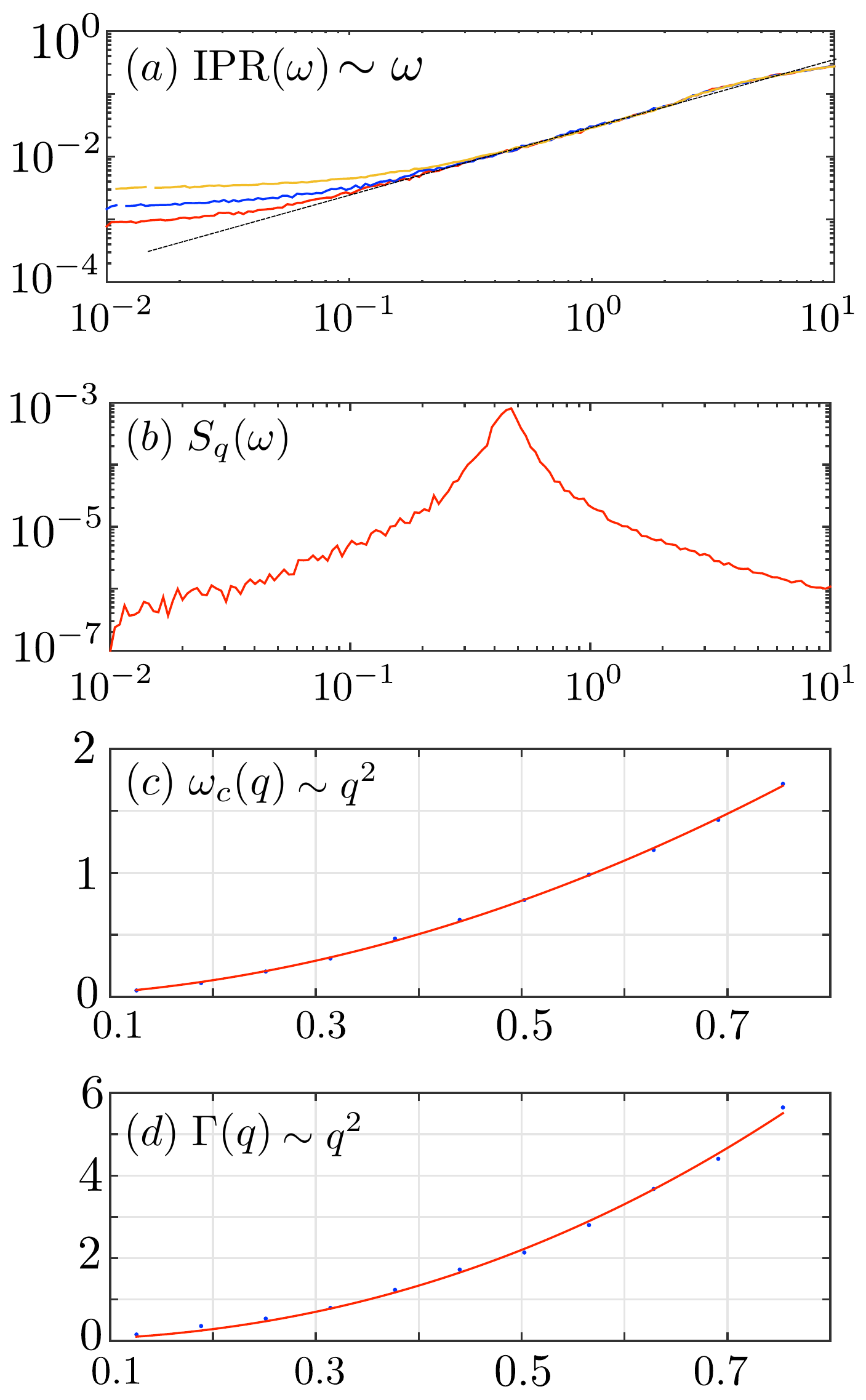}
\caption{(a) The inverse participation ratio is finite [seen by scaling with system sizes $L = 500$ (yellow), $L = 1000$ (blue) and $L = 2000$ (red) ] and scales with frequency as $\text{IPR} (\omega) \sim \omega$ (dashed line fit); (b) spin-wave peak seen in the structure factor $S_q (\omega)$; (c) the peak occurs at frequency $\omega_c (q) \sim q^2$ (red line fit corresponds to $q^{1.95}$); (d) the width (calculated as inverse height of the peak) scales as $\Gamma (q) \sim q^2$ (red line fit corresponds to $q^{2.2}$)}
\label{fig:HPFerro}
\end{figure}
\end{center}

\begin{center}
\begin{figure}[h]
\includegraphics[width=2.5in]{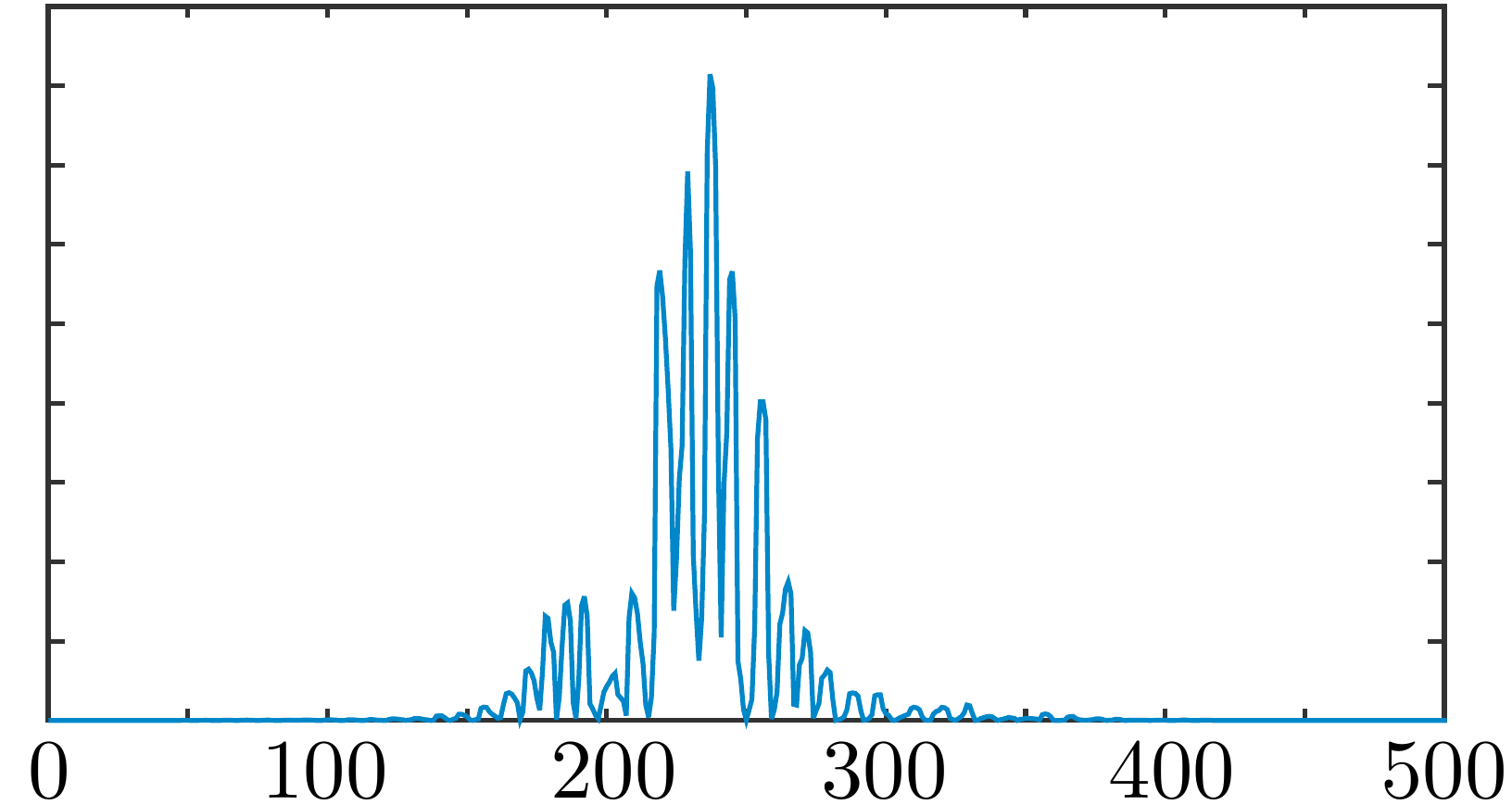}
\caption{An example localized spin-wave wave-function for the disordered purely F chain of size $L = 500$.}
\label{fig:HPFerrowave}
\end{figure}
\end{center}

\begin{center}
\begin{figure}[h]
\includegraphics[width=3in]{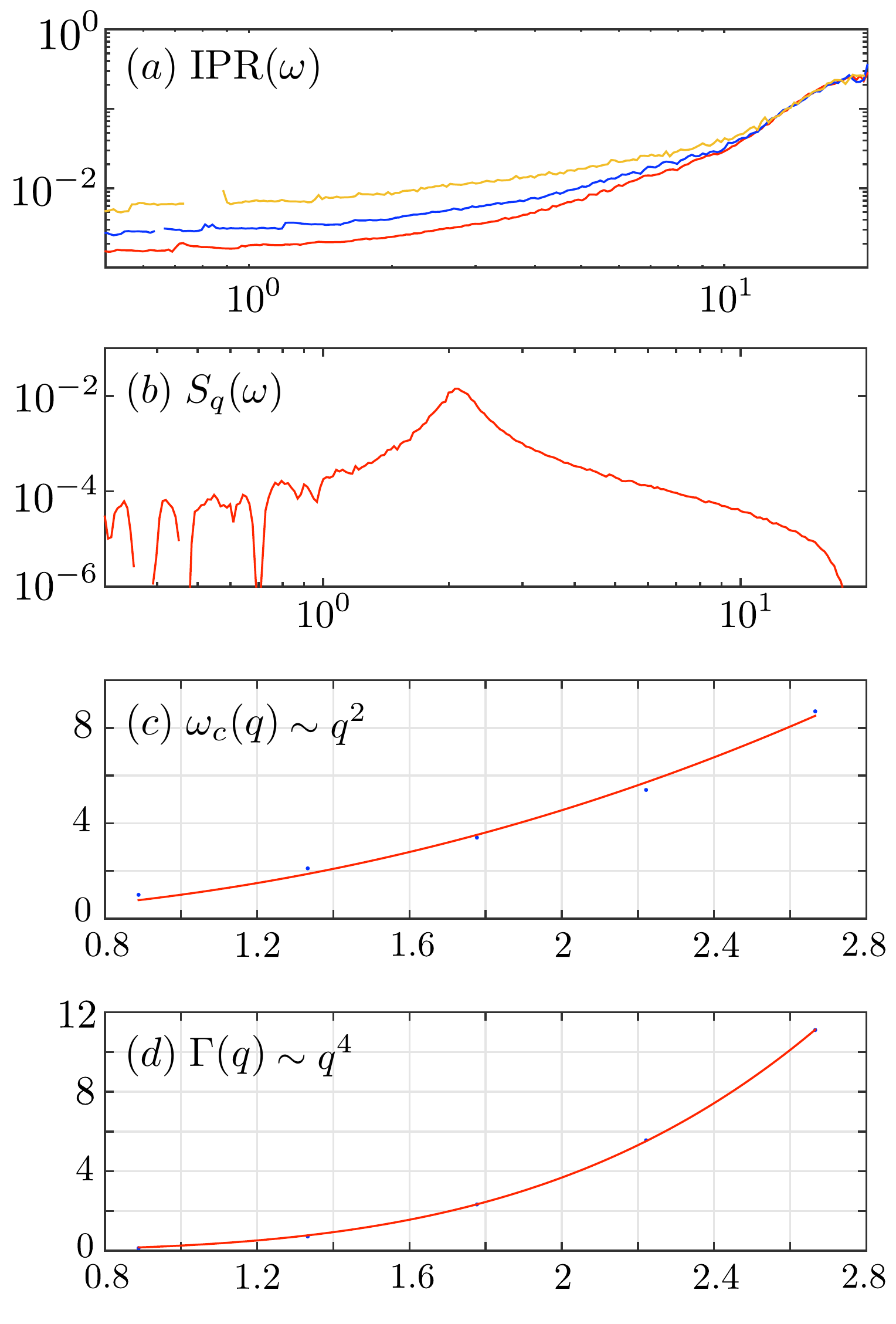}
\caption{(a) The inverse participation ratio decreases to zero in the thermodynamic limit [seen by scaling with system sizes $20 \times 20$ (yellow), $30 \times 30$ (blue) and $40 \times 40$ (red) ] ; (b) spin-wave peak seen in the structure factor $S_q (\omega)$; (c) the peak occurs at frequency $\omega_c (q) \sim q^2$ (red line fit corresponds to $q^{2.1}$); (d) the width (calculated as inverse height of the peak) scales as $\Gamma (q) \sim q^4$ (red line fit corresponds to $q^{3.95}$).}
\label{fig:HPFerro2d}
\end{figure}
\end{center}

\section{Holstein-Primakoff method for $T = 0$}
\label{sec:hpdiffusion}
The RG seems to generally fail for a system that is primarily ferromagnetic. This is indicated by consistently poor fidelity ($f \sim 2$) and a distribution of gaps that fails to develop a singular form. Thus, we need an alternative method to understand the dynamics in the nearly F spin chain. To do so, we analyze the purely F chain by performing a Holstein-Primakoff expansion~\cite{assamagnetism} on its fully polarized ground state. The quadratic Hamiltonian obtained (linearized in the fluctuations) is diagonalized to calculate the structure factor. The simulations were performed for systems of size $L = 500,1000,2000$ with $D = 3$. 

\begin{center}
\begin{figure}[h]
\includegraphics[width=3in]{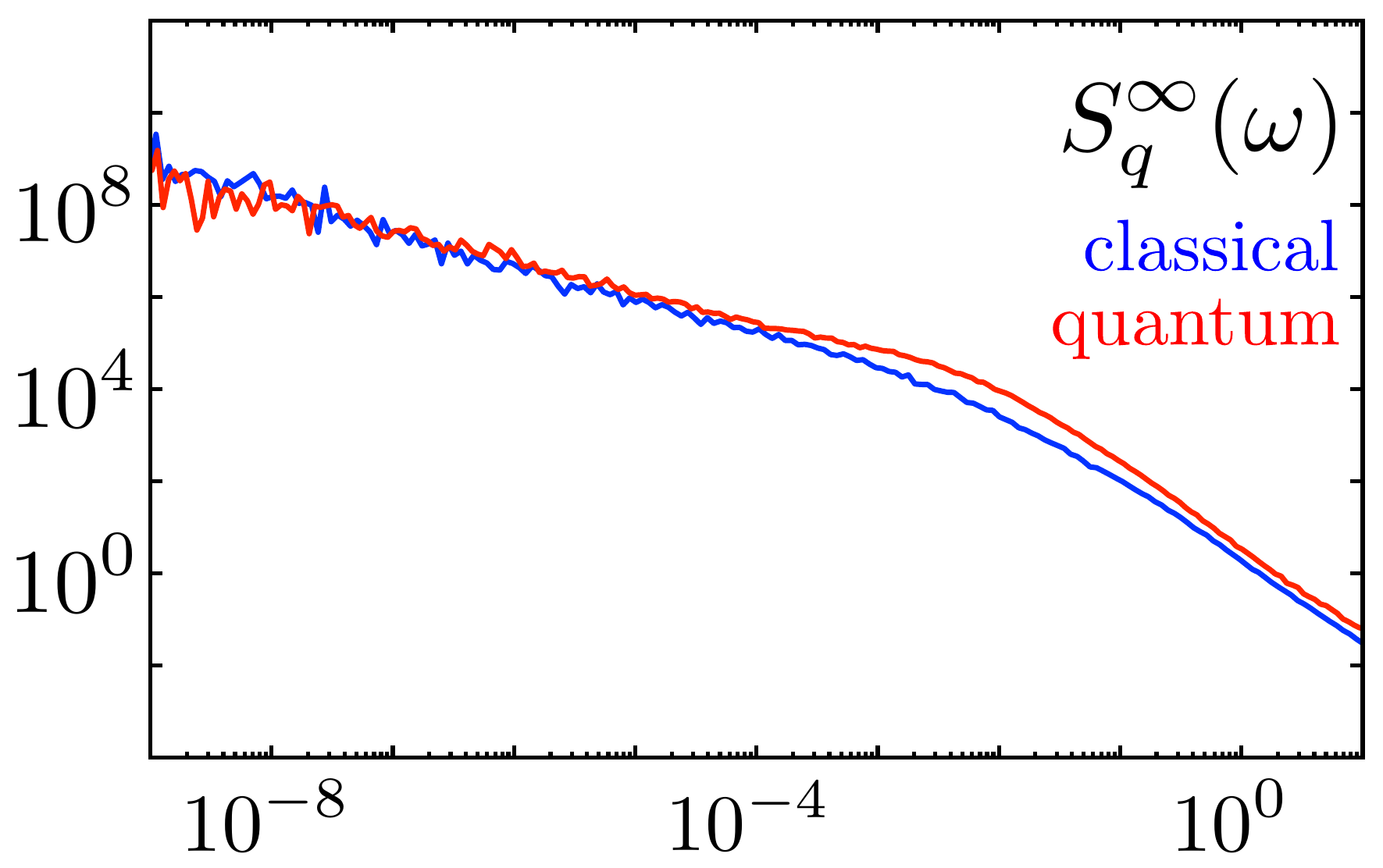}
\caption{Infinite-temperature $S^\infty_q(\omega)$ for classical spins of magnitude $\abs{\vec{S}} = 1/2$ is plotted (blue) and compared with the quantum result (red) at infinite temperature. $L = 15000$, $D = 3$, $\eta_i = 0$.}
\label{fig:classicalquantum}
\end{figure}
\end{center}

The ground state of a fully F spin chain is a classical fully polarized state and the Hamiltonian governing single spin-flip excitations is given by a model of particles in random (Hartree-like) potentials and randomized tunneling terms. Thus, one may expect these excitations to be localized. On the other hand, the failure of the RG suggests that the system is self-averaging and may possess goldstone modes of a disorder-free ferromagnet with a spectral peak at frequency $\omega_c (q) \sim q^2$. It should be noted that the on-site potential and tunneling are exactly correlated in this system, thus, localization is not immediately guaranteed. The results in Fig.~\ref{fig:HPFerro},~\ref{fig:HPFerrowave} suggest that the system exhibits a bit of both; spin-waves exist with a central frequency $\omega_c (q) \sim q^2$, but they are heavily damped with a decay rate $\Gamma$ that scales in the same way, $\Gamma (q) \sim q^2$. Thus, spin waves states are localized in the sense that the inverse participation ratio $\text{IPR}$ [given by $\text{IPR} = \sum_i n_i^2 / (\sum_i n_i)^2$ ; $n_i$ is the spin wave density at site $i$ of the chain] is finite. If one interprets the finite $\text{IPR}$ as a consequence of an exponentially localized wave-function, then $\text{IPR} \sim 1/\xi$, where $\xi$ is the localization length. Extracting the localization length from the $\text{IPR}$ yields the scaling $\xi(\omega) \sim 1/\omega$; this is in agreement with the fact that the RG finds a dynamical exponent $z = 1$ in the limit of the fully F chain which yields a scaling between the cluster size $n$ and the frequency $\omega$ as $n (\omega) \sim 1/\omega^{1/z}$. 

We note that this `localized' spin-wave behavior is unique to one dimension. In two dimensions, an analogous Holstein-Primakoff analysis reveals that the spin waves are sharp; the central frequency $\omega_c \sim q^2$ but the damping scales as $\Gamma \sim q^4$ (see Fig.~\ref{fig:HPFerro2d}). Note that this is in agreement with the analysis of disordered ferromagnets using an effective medium approximation discussed in Refs.~\cite{edwardsferro,yatesferro} (the authors focus on the three-dimensional case, but their results are straightforwardly extended to two dimensions). The simulations for the two-dimensional case were performed for systems of lattice sites $20 \times 20$, $30 \times 30$, $40 \times 40$ and with disorder strength $D = 3$.

\begin{center}
\begin{figure}[h]
\includegraphics[width=3in]{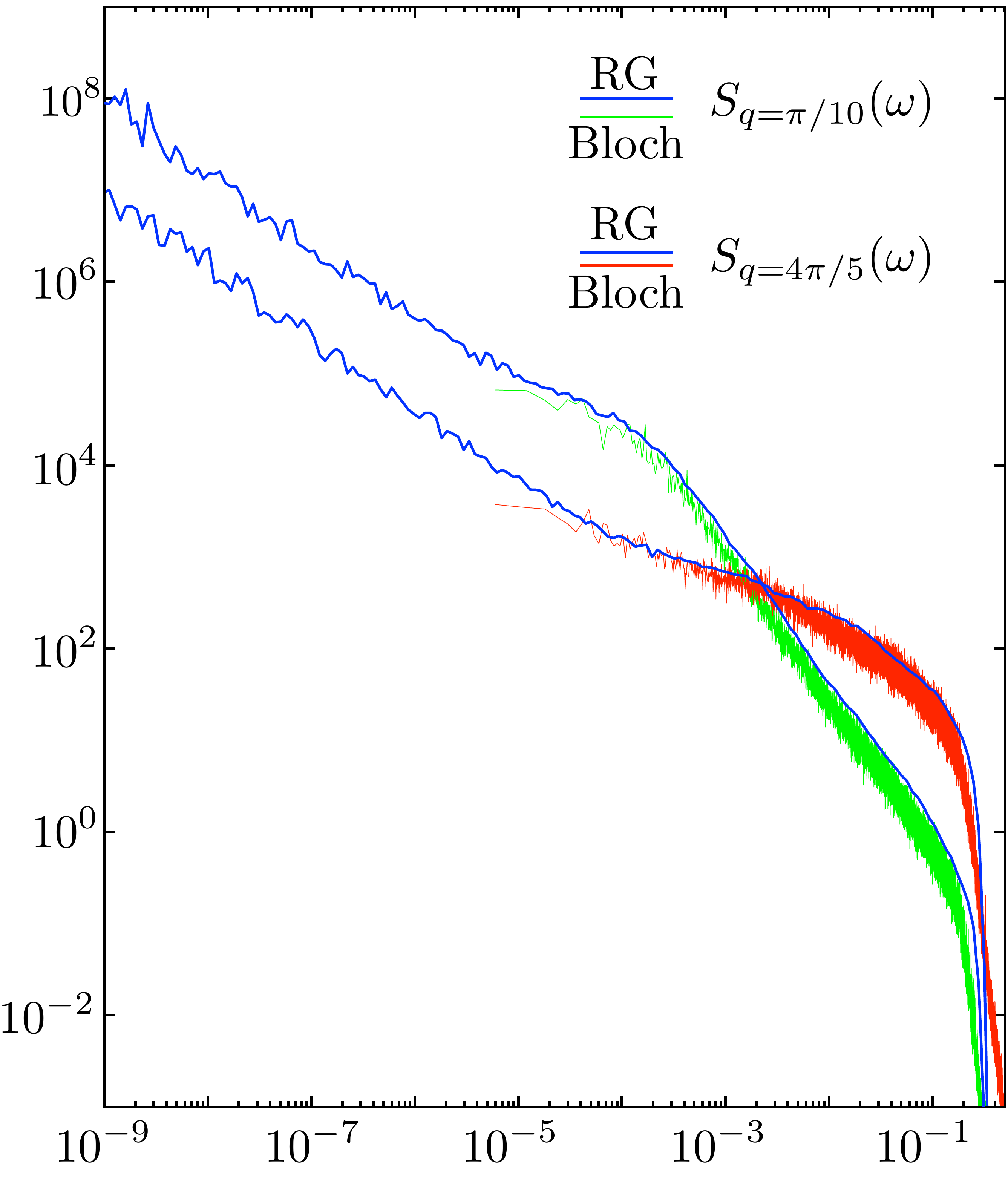}
\caption{Comparison of the structure factor obtained for classical spin chains using direct integration of Bloch-equations and RSRG rules. The direct simulation is computationally hard to perform over very long time-scales, but the agreement is very reasonable for the simulated frequencies.}
\label{fig:classicalcomp}
\end{figure}
\end{center}

\section{Classical simulations for $T = \infty$}
\label{sec:classicalRG}

The RG rules were derived using classical reasoning and can be applied to classical vector spins. Fig.~\ref{fig:classicalquantum} plots (blue) the structure factor at infinite-temperature obtained from such a calculation with classical Heisenberg spins of vector magnitude $\abs{\vec{S}} = 1/2$. The structure factor is very similar to the quantum result (red). We conclude that the anomalous dynamical exponent at infinite temperature, that gives rise to anomalous diffusion is not a purely quantum phenomenon.

\section{Comparison of classical RG with direct dynamical simulation}
\label{sec:rgbloch}

The classical spin chain (size $L = 1000$) with initial coupling distribution $P_0 (|J|) \sim 1/|J|$ with $|J| \in [10^{-3}, 10^{-3} e^5]$ is simulated using the classical Bloch-equations to obtain the structure factor. The result is compared with the classical-RG simulation result in Fig.~\ref{fig:classicalcomp}. The simulations are for $T = \infty$ (the initial spin configuration is chosen entirely randomly).

\end{document}